\documentclass{aa}
\usepackage[varg]{txfonts}
\usepackage{natbib}
\usepackage{graphicx}
\usepackage{float}
\usepackage{lscape}
\usepackage{caption}
\usepackage{subcaption}
\usepackage{sidecap}
\usepackage{xcolor}
\usepackage{comment}
\titlerunning{}
\usepackage{xspace}
\usepackage{amsmath}
\begin{document} 

   \title{The VISCACHA survey -- VIII. Chemical evolution history of the Small Magellanic Cloud west halo clusters}
   \author{Saroon S
          \inst{1}\fnmsep\thanks{saroonsasi19@gmail.com}
          \and
           Bruno Dias\inst{2}
           \and
           Takuji Tsujimoto\inst{3}
           \and
           M. C. Parisi\inst{4,5}
           \and
           Francisco Maia\inst{6}
           \and
           Leandro Kerber\inst{7}
           \and
           Kenji Bekki\inst{8}
           \and
           Dante Minniti\inst{1,9,10} 
           \and
           R. A. P. Oliveira\inst{11}
           \and
           P. Westera\inst{12} 
           \and 
           Orlando J. Katime Santrich\inst{7}
           \and
           Eduardo Bica\inst{13}
           \and
           David Sanmartim\inst{14}
           \and
           Bruno Correa Quint\inst{14}
           \and
           Luciano Fraga\inst{15}
           }
          
   \institute{
   $^{1}$ Instituto de Astrofísica, Facultad de Ciencias Exactas, Universidad Andrés Bello, Fernández Concha 700, Las Condes, Santiago, Chile  \\ 
   $^{2}$ Instituto de Alta Investigaci\'on, Sede Esmeralda, Universidad de Tarapac\'a, Chile\\
   $^{3}$ National Astronomical Observatory of Japan, Mitaka, Tokyo 181-8588, Japan\\
   $^{4}$Observatorio Astronómico, Universidad Nacional de Córdoba, Laprida 854, X5000BGR, Córdoba, Argentina.\\
   $^{5}$Instituto de Astronomía Teórica y Experimental (CONICET-UNC), Laprida 854, X5000BGR, Córdoba, Argentina.\\
   $^{6}$ Instituto de Física - Universidade Federal do Rio de Janeiro, Av. Athos da Silveira Ramos, 149, Rio de Janeiro, 21941-909, Brazil\\
   $^{7}$Universidade Estadual de Santa Cruz (UESC), Departamento de Ci\^encias Exatas, Rodovia Jorge Amado km 16, 45662-900 Ilh\'eus, Brazil\\
   $^{8}$ International Centre for Radio Astronomy Research, The University of Western Australia, 7 Fairway, Crawley, WA 6009, Australia\\
   $^{9}$ Vatican Observatory, V00120 Vatican City State, Italy\\
   $^{10}$Departamento de Fisica, Universidade Federal de Santa Catarina, Trinidade 88040-900, Florianopolis, Brazil\\
   $^{11}$Universidade de São Paulo, IAG, Rua do Matão 1226, Cidade Universitária, São Paulo 05508-900, Brazil\\ 
   $^{12}$Universidade Federal do ABC, Centro de Ciências Naturais e Humanas, Avenida dos Estados, 5001, 09210-580, Brazil\\
   $^{13}$Universidade Federal do Rio Grande do Sul, Departamento de Astronomia,.CP 15051,Porto Alegre 91501-970, Brazil\\
   $^{14}$Rubin Observatory Project Office, 950 N. Cherry Ave., Tucson, AZ 85719, USA\\
   $^{15}$Laborat\'orio Nacional de Astrof\'isica LNA/MCTI, 37504-364 Itajub\'a, MG, Brazil\\
            } 
   \date{Received---- ; accepted----}
  \abstract
   {The chemical evolution history of the Small Magellanic Cloud (SMC) has been a matter of debate for decades. The challenges in understanding the SMC chemical evolution are related to a very slow star formation rate (SFR) combined with bursts triggered by the multiple interactions between the SMC and the Large Magellanic Cloud, a significant ($\sim0.5$~dex) metallicity dispersion for the SMC cluster population younger than about 7.5~Gyr, and multiple chemical evolution models tracing very different paths through the observed age-metallicity relation of the SMC. There is no doubt that these processes were complex. Therefore, a step-by-step strategy is required in order to better understand the SMC chemical evolution. We adopted an existing framework to split the SMC into regions on the sky, and we focus on the west halo in this work, which contains the oldest and most metal-poor stellar populations and is moving away from the SMC, that is, in an opposite motion with respect to the Magellanic Bridge. We present a sample containing $\sim 60$\% of all west halo clusters to represent the region well, and we identify a clear age-metallicity relation with a tight dispersion that exhibits a 0.5~dex metallicity dip about 6~Gyr ago. We ran chemical evolution models and discuss possible scenarios to explain this metallicity dip, the most likely being a major merger accelerating the SFR after the event. This merger should be combined with inefficient internal gas mixing within the SMC and different SFRs in different SMC regions because the same metallicity dip is not seen in the AMR of the SMC combining clusters from all regions. We try to explain the scenario to better understand the SMC chemo-dynamical history.}

\maketitle

\section{Introduction}
The Large Magellanic Cloud (LMC) and the Small Magellanic Cloud (SMC) are an interacting pair of dwarf-irregular galaxies known collectively as the Magellanic Clouds (MCs), and they are probably on their first infall to the Milky Way (MW; see e.g. \citealp{Besla2007,kallivayalil+13}). The LMC and the SMC are respectively located at distances of $49.59 \pm 0.09$~kpc \citep{Pietrzy2019} and $62.44 \pm 0.47$~kpc \citep{Graczyk2020} from Earth. Substructures like the Magellanic Bridge, the Magellanic Stream, and the Leading Arm are signatures of prior interactions between the MCs and with the MW \citep[e.g.][]{nidever+10,Diaz+Bekki2012,donghia+16}.
Previously, these galaxies were believed to have completed multiple orbits around the MW, and hence, the signatures of these interactions were attributed to the effect of the MW on these galaxies \citep[e.g.][]{gardiner+96,Diaz+Bekki2012}. 
Recently, many works have suggested that these galaxies are on their first passage around the MW, such as the revised proper motion estimates from Hubble Space Telescope (HST, \citealp{kallivayalil+13}), simulations (\citealp{Besla2013}), an updated and heavier mass for the LMC than before (\citealp{erkal+19}), and findings on the overdensity of stars in the MW halo (\citealp{conroy+21}). Ultimately, the formation of most observed features could be explained based on the mutual interactions of the MCs prior to approaching the MW.
 
The gravitational disturbances resulting from galaxy-galaxy interactions do have a direct impact on the dynamical evolution and dissolution of the star clusters of the Magellanic System, and its effects are imprinted in their structural, kinematical, and spatial properties \citep[e.g.][]{santos+20,rodriguez+23}. If the star clusters are analysed as a population, they can also show signatures of their past dynamical and chemical evolution history \citep[e.g.][]{nidever+20, PapIV,deBortili2022}.
In particular, there are signatures of merger events that can be understood by studying the chemical evolution of the galaxy. Star cluster systems are amongst the tools to trace the chemical evolution, for example, by determining their age-metallicity relation (AMR). \cite{Tsujimoto&Bekki_2009} proposed that the SMC experienced a major merger where two progenitor galaxies of the SMC with the mass ratios from 1:1 to 1:4 merged at around 7.5~Gyr ago. They fitted chemical evolution models to the AMR from the observational data available at the time and found 
that a dip greater than$0.3$~dex in metallicity was present at around $\sim 7.5$~Gyr. An infall of metal-poor gas was claimed to be the explanation of this observational evidence.

Since 2009, much more data has been collected and more precise and homogeneous analyses have been performed to obtain the AMR of the SMC clusters. The most up-to-date AMR \citep{parisi+22,deBortili2022} shows an intrinsic metallicity dispersion for almost all ages, and the metallicity dip is not so clear anymore. Nevertheless,
\citealp{Dias+14,Dias+16} proposed a framework to understand the SMC chemical evolution puzzle, piece by piece. They classified the SMC cluster population into groups defined as main body, wing and bridge, counter-bridge, and west halo (WH). 
From their Figure 13, \citet[][]{Dias+16} revealed that if we only analyse the WH region, we find a smaller dispersion in the AMR, which makes it possible to solve one piece of the SMC chemical evolution puzzle. Their results for the WH AMR 
agreed with the burst chemical evolution model of \cite{PT98} with lower metallicity dispersion, especially for the younger clusters ($< 6$ Gyr). However, there were two outliers, AM3 and Kron 7, that they tried to explain. If these outliers were confirmed and there were more outliers discovered in the future with similar ages and metallicities, then the major merger model by \citet{Tsujimoto&Bekki_2009} would be favoured, though with the metallicity dip around $\sim$5~Gyr and not $\sim$7.5~Gyr, as in the original work. In fact, some preliminary results from the VIsible Soar photometry of star Clusters in tApii and Coxi HuguA (VISCACHA) survey pointed in this direction \citep{Dias+19Zenodo}.

In this work, we aim to derive the most up-to-date AMR of the SMC WH to constrain one piece of the SMC chemical evolution puzzle. We analyse 15 SMC WH clusters observed in the context of the VISCACHA deep and spatially resolved photometric survey \citep{PapI}, and we combine these data with previous observations from our collaboration to reach a sample of $\sim60\%$ of all WH clusters. We also run chemical evolution models to reproduce our observational results and propose a scenario for the chemical evolution of the SMC.

This paper is organised as follows: In Section 2, we present the data from the VISCACHA survey. The cluster parameters are derived in Section 3. We present the isochrone fitting results in Section 4 and discuss them in the context of the AMR in Section 5. Finally, a summary is presented and conclusions are drawn in Section 6.
\section{The VISCACHA survey}
The VISCACHA survey uses the 4.1 m Southern Astronomical Research (SOAR) telescope, which has a Ground Layer Adaptive Optics (GLAO) module installed in the SOAR Adaptive Optics Module Imager (SAMI; \citealp{tokovini+16}) camera and achieves deep photometry with a spatial resolution of up to $\sim0.3"$ in the $I$ band. The VISCACHA survey has been producing precise CMDs of clusters throughout the LMC, SMC, and the Magellanic Bridge, even for the oldest compact clusters immersed in the dense fields of the MCs, which is often not feasible with large surveys \citep{PapI,dias+20}. The field of view of SAMI is $3.1'\times3.1'$, covering most clusters beyond their tidal radii.
Photometry of these clusters' stars usually reaches the $24^{th}$ magnitude in the $V$ and $I$ filters, thus adequately sampling all the CMD features required to determine the age, metallicity, distance, and reddening of our sample clusters.

The WH region was defined by \citet{Dias+16} on the basis of an ellipse centred on the SMC centre with an eccentricity of $e = 0.87$ and position angle of $PA = 45^\circ$.
There are 41 clusters in the catalogue of \cite{Bica+20} that are part of this region. However, more than half of them are faint and do not have accurate estimates of stellar parameters in the literature before VISCACHA. Our survey obtained the data for 15 clusters during the observation cycles in 2016, 2017, 2019, and 2021 (see observation log in Table \ref{tablog}). 

The bias and flat-field corrections for the data were done following the standard methods using Image Reduction and Analysis Facility (IRAF). Data reduction, Point Spread Function (PSF) photometry, and calibration were performed following the procedures described in \cite{PapI}.
Photometric calibration was carried out using \cite{Stetson_2000} standard star fields. In addition, a few clusters (e.g. Kron\,11) were observed during non-photometric nights and were calibrated using standard coefficients and zero points derived from the Magellanic Cloud Photometric Survey (MCPS; \cite{Zaritsky2002}). 


\section{Cluster parameters}

The $V$, $V-I$ CMDs of the sample clusters display well-defined main sequences (MSs), main sequence turn-offs (MSTOs), and red giant branches (RGBs), see Figs. \ref{Imgdecont} and \ref{ImgdecontCMD}. These features of the CMDs provide essential information about the cluster's global physical properties such as age, metallicity, distance, and reddening. Because there is contamination from field stars, their decontamination is a very important process for studying these objects. Identifying the cluster members using  proper motions from \textit{Gaia} (\citealp{GaiacolabDR3}) is one method, but the \textit{Gaia} magnitude limit is about G $< 21$, which is roughly the magnitude of the MSTO of a $\sim$3~Gyr old SMC cluster and is therefore not suitable for our data (see Figs. \ref{Imgtrial01},\ref{Imgtrial02}, and \ref{Imgtrial03}). 
An alternative method to distinguish the cluster members from the field stars involves using the photometric data by means of statistical comparison of star samples taken from the cluster region and from an offset field, as described in the following section.

\subsection{Statistical decontamination}

\begin{figure*}[t]
\centering
     \begin{subfigure}[b]{0.48\textwidth}
         \includegraphics[width=\textwidth]{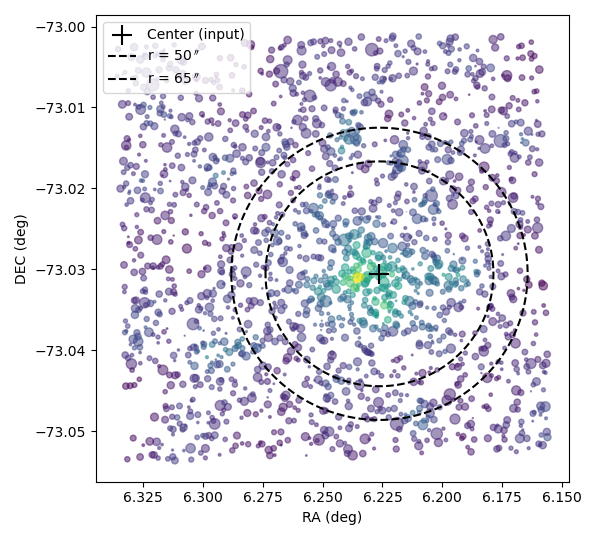}
     \end{subfigure}
      \begin{subfigure}[b]{0.5\textwidth}
         \includegraphics[width=\textwidth]{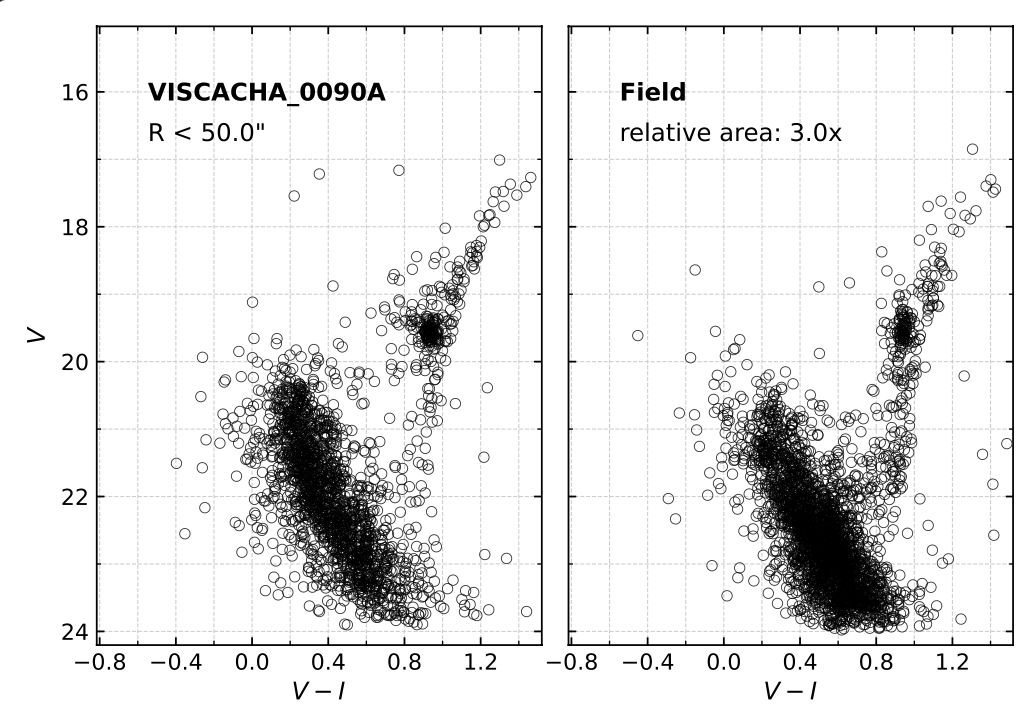}
     \end{subfigure}
    \caption{ Illustration of decontamination procedures on the sample cluster NGC152. {\it Left panel:} Selection of the cluster field and nearby contaminated field for NGC152. The inner radius contains the selected cluster stars. The region outside the outer circle contains the selected field stars. The annular region in between the outer and inner radius is neglected to guarantee that the field is free of possible outer cluster stars. {\it Middle panel:} Corresponding CMDs for the cluster region (within the inner radius). {\it Right panel:} Corresponding contaminated field region (beyond the outer radius). The average grid sizes of $\Delta V = 1$ and $\Delta (V-I) = 0.2$ were adopted to better identify the CMD sequences of this cluster.}
    \label{Imgdecont}
\end{figure*}

Statistical decontamination was carried out following a method adapted from \cite{Maia2010}. The cluster field and the nearby comparison field with similar density and reddening were selected. The selection of cluster population and field stars for the sample cluster NGC152 is illustrated in Figure \ref{Imgdecont}.  The cluster's 
radius was 
adopted from the catalogue of \cite{Bica+20}. 
The CMDs for the selected cluster and field stars were binned into cells small enough to detect the local variation of the field-star contamination on the various sequences in the CMD but still contain a significant number of stars. The average cell sizes were derived for each cluster from the dispersion of the data across each magnitude and colour, yielding typical values of $\Delta V = 0.75$ mag and $\Delta (V-I) = 0.33$ mag, respectively. The membership probability was assigned based on the overdensities of the cluster stars in relation to the field stars. 
This method has been extensively used in previous VISCACHA works as well as in the analysis of Galactic (\citealp{Angelo+18}) and MC clusters (\citealp{Maia+2014}). Additional details can be found in \citet{Maia2010}.

\begin{figure*}[t]
\centering
     \begin{subfigure}[b]{0.33\textwidth}
         \includegraphics[width=\textwidth]{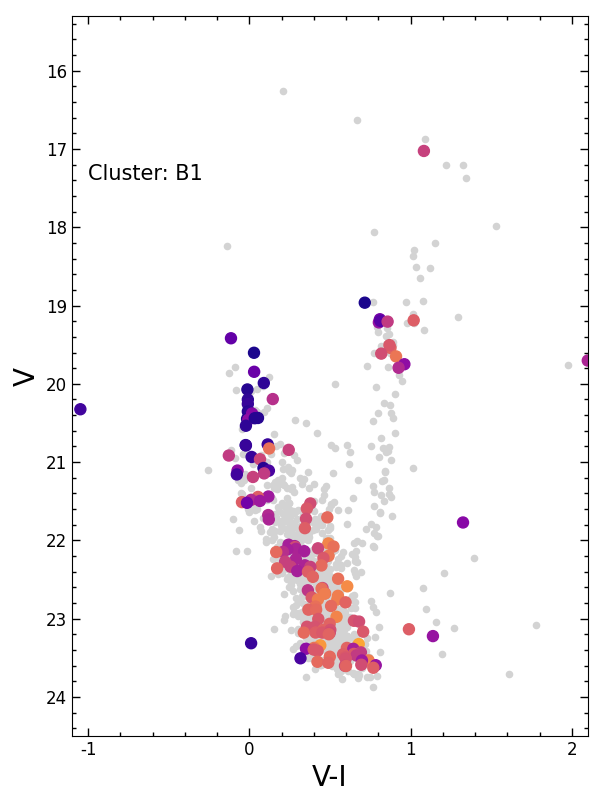}
     \end{subfigure}
      \begin{subfigure}[b]{0.33\textwidth}
         \includegraphics[width=\textwidth]{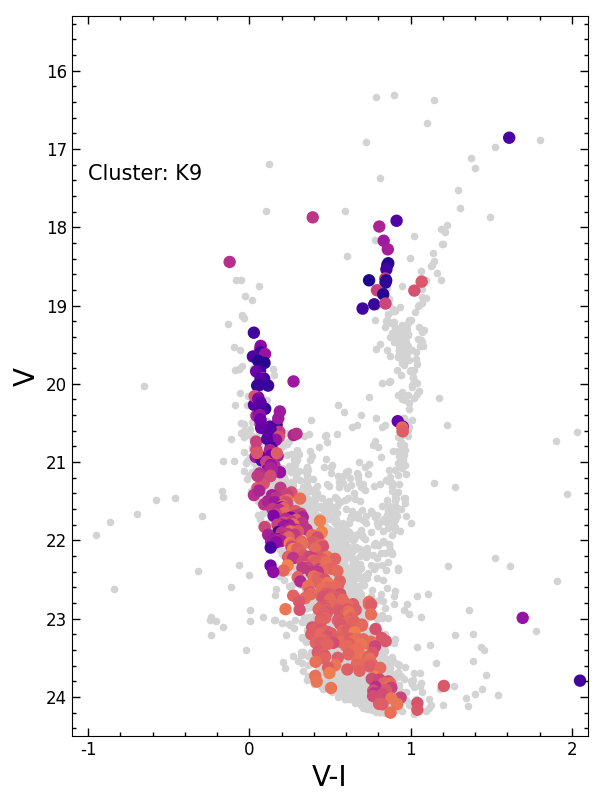}
     \end{subfigure}
     \begin{subfigure}[b]{0.33\textwidth}
         \includegraphics[width=\textwidth]{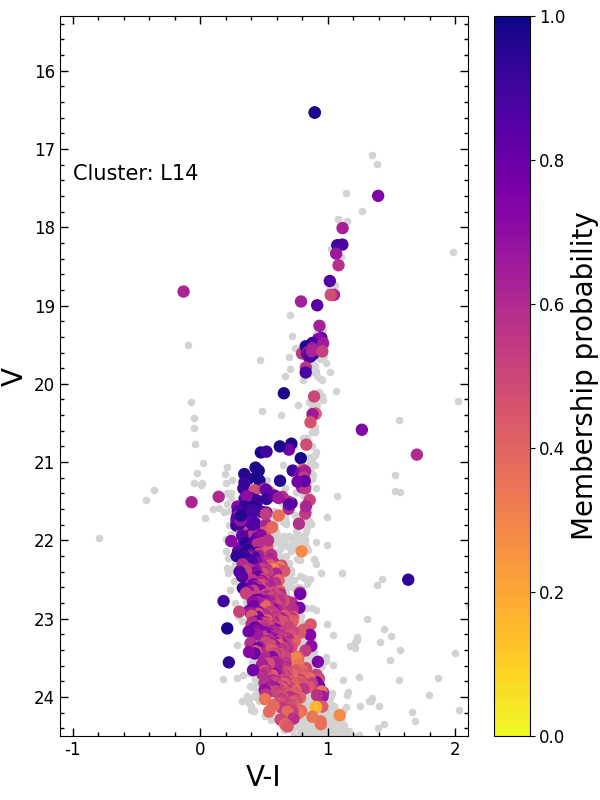}
     \end{subfigure}
    \caption{
    Decontaminated CMDs for three example clusters from our sample. 
    The coloured points are the stars belonging to the cluster population. The grey points represent the field stars. The colours represents the membership probability of the cluster population. }
    \label{ImgdecontCMD}
\end{figure*}

Figure \ref{ImgdecontCMD} shows the decontaminated CMDs of three sample clusters. The cluster B1 is analysed for the first time in this work. The cluster K9 is much younger than previously thought. The cluster L14 is one of the oldest and most massive clusters in our sample. In the figure, the removed field stars are indicated by grey points, and the selected cluster stars are indicated by colours, which also represent the membership probability of each star.

\subsection{Isochrone fitting}
%
In our study, the statistically decontaminated CMD of each cluster was fitted with PARSEC isochrones \citep{Bressen2012} in order to derive more accurate determinations of age, metallicity, distance, and reddening.
The so-derived astrophysical parameters for all the clusters are given in Table \ref{tabres}. 
The uncertainties were obtained by fitting isochrones that wrap the boundaries of the CMD features and are still reasonable fits. These isochrones are shown as dash-dotted lines in Figs. \ref{Imgtrial01}, \ref{Imgtrial02}, and \ref{Imgtrial03}. 
The derived uncertainties are also listed in Table \ref{tabres}.

\begin{table*}
    \centering
    \caption{Astrophysical parameters derived from the isochrone fits.}
    \label{tabres}

    \begin{tabular}{p{3cm}p{3cm}p{3cm}p{3cm}p{3cm}}
         \hline
         \noalign{\smallskip}
         Cluster Name & Age & [Fe/H] & Dist  & $A_v$  \\
         & [Gyr] &&[kpc]& [mag]\\ 
        \noalign{\smallskip}
         \hline\hline
        \noalign{\smallskip}
         
         K9         &$0.80_{-0.10}^{+0.10}$      &$-0.90_{-0.05}^{+0.05}$       & $54.0_{-1.0}^{+1.0}$       & $0.22_{-0.12}^{+0.10}$   \\
         \noalign{\smallskip}
         B1         &$0.90_{-0.10}^{+0.10}$      &$-0.80_{-0.10}^{+0.10}$       & $67.5_{-1.5}^{+2.5}$       & $0.15_{-0.14}^{+0.10}$   \\
         \noalign{\smallskip}
         NGC152     &$1.30_{-0.10}^{+0.10}$      &$-0.80_{-0.10}^{+0.20}$       & $60.0_{-2.0}^{+4.0}$       & $0.31_{-0.11}^{+0.09}$   \\
         \noalign{\smallskip}
         B2         &$1.60_{-0.10}^{+0.20}$      &$-0.90_{-0.10}^{+0.20}$       & $62.5_{-4.5}^{+1.5}$       & $0.34_{-0.14}^{+0.06}$   \\
         \noalign{\smallskip}
         K6         &$1.70_{-0.10}^{+0.10}$      &$-0.70_{-0.05}^{+0.10}$       & $61.0_{-2.0}^{+2.0}$       & $0.28_{-0.08}^{+0.08}$   \\
         \noalign{\smallskip}
         K8         &$2.50_{-0.10}^{+0.20}$      &$-0.80_{-0.10}^{+0.10}$       & $64.0_{-2.0}^{+2.0}$       & $0.10_{-0.09}^{+0.08}$   \\
         \noalign{\smallskip}
         B4         &$2.75_{-0.25}^{+0.05}$      &$-0.98_{-0.12}^{+0.13}$       & $57.0_{-2.0}^{+8.0}$       & $0.30_{-0.10}^{+0.10}$   \\
         \noalign{\smallskip}
         K7         &$2.65_{-0.20}^{+0.05}$      &$-0.95_{-0.05}^{+0.05}$       & $65.5_{-1.5}^{+1.5}$       & $0.15_{-0.07}^{+0.07}$   \\
         \noalign{\smallskip}
         L14        &$3.20_{-0.40}^{+0.20}$      &$-1.00_{-0.10}^{+0.10}$       & $63.0_{-2.0}^{+4.0}$       & $0.12_{-0.09}^{+0.09}$   \\
         \noalign{\smallskip}
         K11        &$3.40_{-0.20}^{+0.60}$      &$-1.00_{-0.20}^{+0.10}$       & $56.7_{-2.7}^{+1.3}$       & $0.27_{-0.17}^{+0.13}$   \\
         \noalign{\smallskip}
         L2         &$4.00_{-0.15}^{+0.20}$      &$-1.30_{-0.20}^{+0.10}$       & $56.5_{-1.5}^{+1.5}$       & $0.32_{-0.22}^{+0.12}$   \\
         \noalign{\smallskip}
         HW1        &$4.60_{-0.10}^{+0.10}$      &$-1.20_{-0.10}^{+0.10}$       & $62.0_{-4.0}^{+4.0}$       & $0.18_{-0.12}^{+0.13}$   \\
         \noalign{\smallskip}
         AM3        &$4.70_{-0.30}^{+0.60}$      &$-1.10_{-0.10}^{+0.10}$       & $64.8_{-3.8}^{+3.2}$       & $0.11_{-0.06}^{+0.08}$   \\
         \noalign{\smallskip}
         HW5        &$4.90_{-0.40}^{+0.40}$      &$-1.30_{-0.10}^{+0.10}$       & $62.0_{-4.0}^{+4.0}$       & $0.29_{-0.14}^{+0.11}$   \\
         \noalign{\smallskip}
         NGC121     &$10.1_{-0.50}^{+0.30}$      &$-1.50_{-0.15}^{+0.05}$       & $68.0_{-2.0}^{+2.0}$       & $0.17_{-0.08}^{+0.13}$   \\
         
         \noalign{\smallskip}  
         \hline
    \end{tabular}

\end{table*}

Multi-band photometry sometimes helps constrain the metallicities of the star clusters ( e.g. by checking the slope of the RGB and the location of RC stars). For this reason, we used complementary data from other surveys even though they present shallower photometry or seeing-limited photometry that does not reach fainter stars in the cluster cores.
We chose the surveys \textit{Gaia DR3} (\citealp{GaiacolabDR3}), the Survey of the Magellanic Stellar History (SMASH DR2; \citealp{SMASHsurvey}), and the VISTA survey of the Magellanic Clouds system (VMC DR5; \citealp{VMCsurvey}), as they contain observational data of our targets under different conditions and using different filters. The catalogues from each of these surveys were matched with the data from our VISCACHA survey. The membership probabilities of the stars from the decontaminated VISCACHA samples were assigned to the matched stars in the corresponding clusters from all the aforementioned surveys. The CMDs for each of these surveys were plotted in their respective filters, and PARSEC isochrones were fitted using the best fitting parameters obtained from VISCACHA CMDs. Although Gaia and VMC CMDs are not deep enough to present a well-defined MS and MSTO, they were very useful in verifying the RGB slope and the fit of the isochrone with respect to the RC location. Using the RC location, we could establish a position in the RGB and identify it evenly in multi-band CMDs. To expand the wavelength coverage, the CMDs in $V-Ks, V$ were plotted using the combined VISCACHA and VMC filters. We employed the extinction law by \cite{cardelli1989relationship} to consistently correct the PARSEC isochrones as indicated in the PARSEC online form. The best-fit plots for all the clusters, including the multi-band CMDs, are presented in the Appendix in Figures \ref{Imgtrial01}, \ref{Imgtrial02}, and \ref{Imgtrial03}. 
The grey points in the figures represent the field stars.
The coloured points represent the most probable cluster
population. The solid lines show the best-fit isochrones, and the dash-dotted lines represent the uncertainties. 
We found it impressive that the isochrone fitting to the VISCACHA CMD translates into a good match between isochrone and data from other surveys with different filters and wavelength ranges. This multi-band analysis gives extra strength to our results.

\section{Results}

\begin{figure*}[t]
    \centering
    \begin{subfigure}[b]{0.45\textwidth}
         \includegraphics[width=\textwidth]{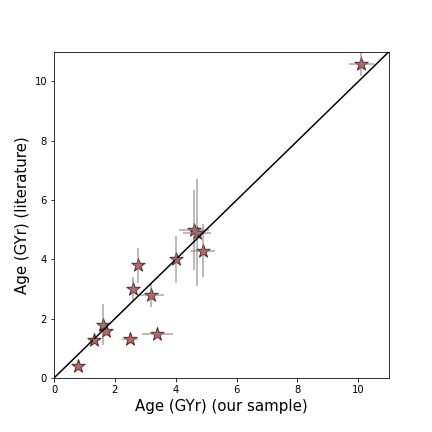}
     \end{subfigure}
      \begin{subfigure}[b]{0.45\textwidth}
         \includegraphics[width=\textwidth]{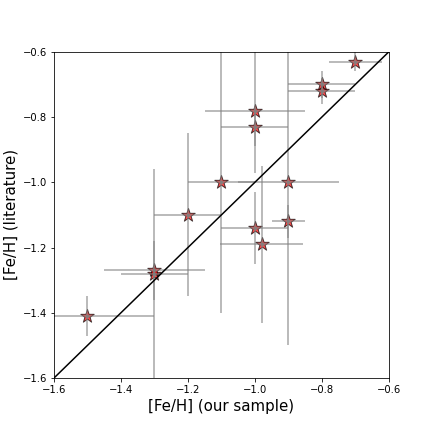}
     \end{subfigure}
    \caption{Correlation plots illustrating the comparison between the derived ages (and metallicities) with best representing ages (and metallicities) obtained from literature for 14 of the clusters in common from our sample.    }
    \label{ImgAMCorltn}
\end{figure*}

We fit PARSEC isochrones to the decontaminated cluster CMDs using VISCACHA data while varying all four parameters of age, metallicity, distance, and extinction simultaneously
with homogeneous data (i.e. we did not fix metallicity or distance from other sources to perform the fit). We found a very good fit for all 15 clusters, as displayed in Figs. \ref{Imgtrial01}, \ref{Imgtrial02}, and \ref{Imgtrial03}. In particular, the cluster $Br\ddot uck$ 1 (B1) was analysed for the first time ever. Table \ref{tabres} provides the astrophysical parameters and uncertainties derived for the 15 sample clusters in the SMC WH region. Even though the uncertainties were derived using a visual procedure, the values overall are of the same order as the uncertainties derived by statistical analysis in previous VISCACHA papers, and they are realistic, as is shown by the dot-dashed line isochrones wrapping most of the cluster CMD stars in Figs. \ref{Imgtrial01}, \ref{Imgtrial02}, and \ref{Imgtrial03}. The isochrones corresponding to the same parameters were plotted over the multi-band CMDs generated from other surveys (i.e. Gaia, VMC, and SMASH) in order to check whether the comparison between data and PARSEC models is still consistent at other wavelengths. In most cases, the match is acceptable.
The only exceptions are RGB stars with combined colour between VISCACHA and VMC for the clusters B\,4 and L\,14, as the B\,4 RGB stars tend to be bluer (more metal poor) and the L\,14 RGB stars tend to be redder (more metal rich) than the isochrones. For these two clusters, we have preliminary spectroscopic metallicities of [Fe/H] = $-0.99\pm0.05$ and $-1.16\pm0.05$ ( Parisi et al. in prep.), respectively, which are in good agreement with the photometric metallicities derived from the VISCACHA CMD (i.e. [Fe/H] = $-0.98_{-0.12}^{+0.13}$ and $-1.00\pm0.10$), meaning that the good isochrone fit to the VISCACHA CMD adequately represents the global parameters of B\,4 and L\,14. Therefore, the mismatch between the model and isochrone may be related to other issues intrinsically connected to the limited knowledge on stellar evolution imprinted in the isochrones or systematic differences between the photometric zero points in different wavelengths and different surveys.


 Most of the clusters in our sample (14 out of 15) were previously studied by different authors, as listed in Table \ref{tabcomp}. However, many previous studies had some limitations. For example, some works derived only the ages or only the metallicities of the clusters. Others had age estimates that were derived adopting the mean SMC values for distances and metallicities, inducing systematic errors. Only in \cite{Dias+14,Dias+16,PapIV} was the field star contamination removed. In some cases, the photometry was shallower and had larger errors. As we needed the largest possible sample of ages and metallicities from WH clusters in order to constrain the AMR, this meant that we had to increase our sample by joining it with results from the literature. Therefore, we used the 14 clusters in our sample that have ages and metallicities from the literature to discuss how we can combine our results with those based on other clusters analysed in the literature. 
The results from previous studies show good correlation with the homogeneously derived ages and metallicities from our work. This one-to-one relation between the ages and metallicities is illustrated in Fig. \ref{ImgAMCorltn}. 
 We scrutinise the literature results in the following paragraphs below, cluster by cluster, in light of the homogeneous data and analysis from this work with the derived age, metallicity, distance, and extinction.


We provide the astrophysical parameters for B1 (${\rm Br\ddot uck}$ 1) for the first time. The best fitting isochrone yields the age 0.9 Gyr, [Fe/H] = -0.80, d = 67.5 kpc, and $A_v = 0.15$ mag. The isochrones with similar age and metallicity show excellent fits with the corresponding CMDs using the \textit{Gaia}, VMC, and SMASH filters.

Our best fitting isochrone for the Lindsay 14 cluster resulted in an age of $3.20$ Gyr, [Fe/H]$= -1.00$, $d = 63.0$ kpc, and $A_v = 0.12$ mag. The metallicity is supported by the preliminary result from CaII triplet (CaT) spectroscopy, that is, [Fe/H]$= -1.16$ dex (Dias et al. in prep.). If we fix this exact metallicity and fit the other three parameters to our data, we find 
an age corresponding to 3.5 Gyr, d = 64.5 kpc, $A_v = 0.15$ mag, and the corresponding isochrone fit to be in good agreement with our original fit without any prior, which gives us confidence in our result. The metallicity estimated by \cite{Dias+16} is also consistent with the two results above. However, they found a larger distance (70.6 kpc) and a younger age (2.8 Gyr). This difference can be due to the filters they used and, in particular, to the higher quality and deeper photometry available in the present work, which should have more accurate parameters.

The case of ${\rm Br\ddot uck}$ 4 is similar to Lindsay\,14. Our best fitting isochrone for this cluster resulted in an age of $2.75$ Gyr, [Fe/H]$= -0.98$, $d = 57.0$ kpc, and $A_v = 0.30$ mag, and the metallicity is in excellent agreement with the preliminary result from CaT spectroscopy, that is, [Fe/H]$= -0.99$ (Dias et al. in prep.). \cite{Dias+16} found an older age (3.8 Gyr), a larger distance (66.6kpc), and a somewhat compatible metallicity (-1.19 dex) based on data with lower quality and shallower photometry.

The NGC\,121  cluster's tidal radius is larger than the field of view (FOV) of SAMI, which means that we do not have an adjacent field to perform a statistical decontamination, as was done for the other clusters. Nevertheless, the cluster has so many stars that the CMD is dominated by cluster stars, which made it possible to fit an isochrone to the highest density CMD features.
The central region is so crowded that the confusion limit was reached, and we decided to avoid this region. We restricted the CMD to the outskirts where the photometry is more reliable and still dominated by the cluster. The CMD for radius > 70" is well defined, and the isochrone fitting yielded the parameters $age = 10.10$ Gyr, [Fe/H]$= -1.5$ dex, $d = 68.0$ kpc, and $A_v = 0.17$ mag (see Fig. \ref{Imgtrial03}). Using high-resolution spectroscopy, \cite{Dalessandro16} derived a metallicity of $-1.28 \pm 0.06$ dex, and the preliminary analysis by \cite{Johnson2004} yielded a value of [Fe/H] $= -1.41 \pm 0.14$. The result using low-resolution spectroscopy from \cite{Dacosta1998} is -1.19 dex in the scale defined by \cite{CG97}, which becomes -1.36 dex in the updated \cite{Caretta2009} scale\footnote{The \cite{Caretta2009} scale is consistent with the most up-to-date metallicity scale by \cite{Dias16b} for metallicities up to -0.7. Therefore, the \cite{Caretta2009} scale is good enough in this case.} and is in excellent agreement with the aforementioned high-resolution spectroscopic metallicities. Our photometric metallicity of $-1.5$ is only 0.15 dex lower and is in good agreement with spectroscopy. We note that \citet{Glatt2008b} adopted [Fe/H]$=-1.5$ to derive the age $10.5\pm0.5$ Gyr for this cluster by fitting an isochrone to HST photometry,
which is in excellent agreement with our results.

  \cite{Dias+16} observed the cluster Kron 11 in a bad seeing, and the resulting CMD had shallower photometry that does not define the cluster MS well. According to their findings, Kron 11 is younger (1.47 Gyr) and more metal rich ($-0.78$) than what we found (3.4 Gyr, $-$1.0), though our results are based on a well-defined and deeper CMD.

${\rm Br\ddot uck}$ 2 and HW\,5 were also analysed by \cite{Dias+16}. Their derived ages and metallicities for the clusters are in good accordance with our estimates. 

Kron\,9 is the youngest cluster in our sample. \cite{Glatt2010} and \cite{Nayak2018} derived an age of 0.5 Gyr and 0.4 Gyr, respectively, based on the shallower photometric surveys MCPS and OGLE, whereas we derived 0.8~Gyr. The younger ages were ruled out with our photometry. The metallicity derived by our isochrone fit together with the age of 0.8~Gyr is [Fe/H]$=-0.90$ dex, 
which is 0.2~dex more metal rich than the CaT spectroscopic value of $-1.12$~dex (\citealp{Parisi2015}). However, the CaII technique requires a good determination of the red clump magnitude level, which is trivial for clusters older than about 1.5 Gyr \citep[e.g.][]{Dias+Parisi2020}. Kron\,9 is immersed in a populous field dominated by older stars with a clear red clump $\sim0.4$ mag fainter than that of the cluster, as indicated by our decontaminated photometry and isochrone fit (see Fig. \ref{Imgtrial01}). Without this new information, \cite{Parisi2015} assumed the red clump level to be closer to the field stars red clump, which shifted their metallicity down by $\sim 0.11$ dex. In conclusion, the corrected CaT metallicity is $-1.0$ dex, which is only 0.1 dex different from our photometric metallicity. In other words, they are compatible.

The Kron\,6 system has a metallicity estimate derived from the CaT spectroscopy by \cite{Parisi2015}, which is reported to be -0.63 dex. This metallicity estimate is consistent with the metallicity of -0.70 dex derived from our best fitting isochrone. Additionally, the age of 1.70 Gyr for Kron 6 that was derived together with the -0.70 dex metallicity is in excellent agreement with the age estimate of 1.60 Gyr reported by \cite{2011Piatti}.

 Until this work, HW\,1 was analysed only by \cite{Dias+14} using shallower photometry. Nevertheless, their results (5 Gyr, $-1.10$ dex) are in reasonable agreement with ours (age = 4.60 Gyr and [Fe/H] $= - 1.20$). However, while checking for all possible solutions, we found another reasonable fit particularly for this cluster with a more metal-rich isochrone (4.5 Gyr, $-0.90$ dex). This second set of parameters does not affect our conclusions despite increasing the dispersion.

The astrophysical parameters for the remaining five clusters (namely, NGC152, Kron 8, Kron 7, Lindsay 2, and AM3)\ have been studied by \cite{PapIV} using the same VISCACHA survey data used in this work, but with a slightly different technique. We re-analysed them with the techniques of this work in order to be as homogeneous as possible. The results are in very good agreement, as expected, and we discuss them case by case in the following paragraphs.

As discussed in \cite{PapIV}, the cluster NGC\,152 presents an extended MSTO (eMSTO), and the posterior distribution of their statistical isochrone fitting shows that metallicities range from $\sim -0.7$ to $ -0.9 $. \cite{Crowl2001} presented a more metal-poor value of $ -0.94\pm0.15$~dex, determined from the RGB slope calibration that has larger uncertainties. Our photometric metallicity of $-0.80\pm0.10$~dex is in agreement with the spectroscopic metallicities derived from the CaT (\citealp{PapIV,parisi+22}). The metallicity estimated by \cite{Song+2021} using high-resolution spectroscopy also shows excellent agreement with our estimates.

In the case of the cluster Kron\,8, CaT metallicities of $-0.75\pm0.07$ (\citealp{PapIV}) and $-0.70\pm0.03$ (\citealp{Parisi2015}) are in good agreement with our measurement of $-0.80\pm0.10$. The CMD fitting in \cite{Dias+14} based on shallower photometry gives a more metal-poor result ($-1.12\pm0.24$) with larger uncertainties. 

The Lindsay\,2 cluster has previously been analysed by \cite{Dias+14} and \cite{Dias+16} using isochrone fitting, and their derived ages are in excellent agreement with our estimated age of 4 Gyr. Our best fitting also yields a metallicity of -1.3 ± 0.10, which shows excellent agreement with the CaT metallicity (-1.28) derived in \cite{PapIV}.

The cluster AM\,3 has been studied before using various techniques, as described in \cite{dacosta99,2011Piatti, Dias+14, PapI, PapIV}, resulting in essentially two groups of results: one older and more metal poor ($\sim$5-6 Gyr and $\sim-1.3$) and another younger and more metal rich ($\sim$4.5-5 Gyr and $\sim-0.8,-1.0$). The differences are related to an age-metallicity degeneracy that was difficult to break because the cluster has only a few stars in the RGB. Deeper photometry combined with CaT spectroscopy gave a stronger constraint for the results from \cite{PapIV}. In our analysis, we obtained a metallicity of $-1.10\pm0.10$, which is in good agreement with the CaT metallicity of $-1.00\pm0.10$ from \cite{PapIV}. The ages derived from \citealp{PapIV} (4.40 Gyr) and \citealp{Dias+14} (4.90 Gyr) also agree well with our estimates.

\begin{figure*}[t]
    \centering
      \begin{subfigure}[b]{\textwidth}
         \includegraphics[width=\textwidth]{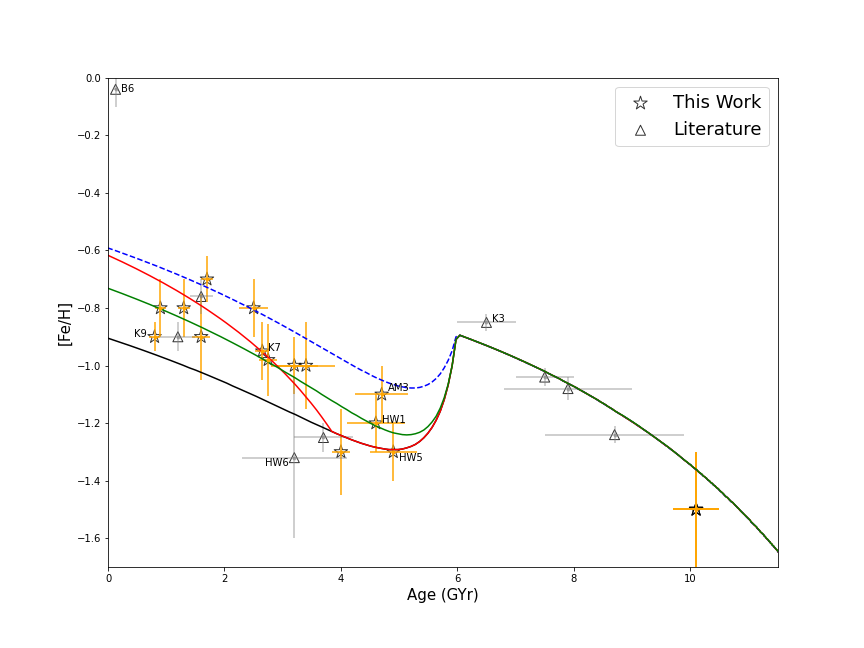}
     \end{subfigure}
    \caption{Age-metallicity relation for SMC WH star clusters. Stars denote the results from this work, and triangles indicate the results from the extended sample from the literature (see text and Tables \ref{tabres} and \ref{tabcomp} for more details). The extended curves represent the results of the model in which the merger has mass ratios of 1:4 (black solid curve) and 1:2 (blue dashed curve). The green solid curve represents the merger model with a 1:4 mass ratio that took place 6~Gyr ago and caused an enhanced SFR, whereas the red solid curve is the scenario with an enhanced SFR 2~Gyr later. }
     
    
    \label{Imgamr}
\end{figure*}

 The cluster Kron\,7 was previously studied using CaT spectroscopy by \citealp{PapIV} and \citealp{parisi+22}. However, there were differences in the selection of members, with a difference in mean metallicity of about 0.3 dex and mean radial velocity (RV) of about 19km/s. We investigated this discrepancy and found that by applying an offset of 16.9km/s to the RVs of \citet{parisi+22} and using \citet{Song+2021} as a reference, the RV of Kron 7 becomes compatible with what was derived by \citet{PapIV}. Moreover, \citet{PapIV} has only one cluster in common with \citet{Song+2021}, and both present a similar RV. We found a common star between the two samples by \citet{PapIV} and \citet{parisi+22}, with an offset of 17.5km/s, and we applied this RV offset to all Kron 7 stars analysed by \citet{parisi+22}. This single common star presents no offset in metallicity (a negligible difference of 0.03 dex), as expected based on other comparisons between NGC152 and Kron8 that were also in common between these two works (see \citealp{PapIV} and \citealp{deBortili2022} for details). We joined both samples for Kron\,7 after putting the RV to the same scale and then re-assessed the membership analysis. We updated the mean RV and metallicity for Kron\,7 to be $144.4\pm1.3$~km/s and $-0.94\pm0.03$~dex, which is in very good agreement with the metallicity found in this work (-0.95 dex), once again supporting the general agreement between our photometric metallicities and those derived with CaT spectroscopic analysis.

\section{The chemical evolution of the Small Magellanic Cloud west halo}
\subsection{The age-metallicity relation}

In addition to the 15 clusters analysed in this work that show good agreement with CaT metallicities and HST-based ages as discussed in the previous section, there are ten more WH clusters with ages and metallicities from the literature by \cite{Parisi2009,Parisi2015,parisi+22}, \citealp{Dias+16}, \citealp{PapIV}, and references therein. We joined our results with this additional sample, ending up with 25 WH clusters that represent well the 41 clusters catalogued by \citet{Bica+20} in this region. This sample displays a clear chemical enhancement at the early times of the SMC life, between 10 to 6 Gyr ago reaching $\sim-0.8$. At the end of this period, something happened and the metallicity sharply decreased by $\sim 0.5$. After this, the chemical enhancement resumed until reaching $\sim-0.8$ once again about $\sim 1$~Gyr ago (see Fig. \ref{Imgamr}).

We discuss each points in the AMR case by case in the following paragraphs. We argue that this combined sample can be analysed as a quasi-homogeneous sample, allowing a reliable AMR for the SMC WH to be traced.


${\rm Br\ddot uck}$ 6 (B6) is the prominent outlier in the AMR. The parameters derived by \cite{Dias+16} show that this cluster is extremely young (0.13 Gyr) and metal rich ([Fe/H] $= -0.04$) compared to the other WH clusters. We checked Gaia DR3 proper motions for this cluster and found that they are different from the average motion of the WH (as reported by \citealp{PapIV}). Therefore, we speculate that B6 is not part of the chemical evolution history of its current neighbours, and today, it only happens to be at this position by coincidence.

Kron 3 is at the metal-rich tip of the AMR curve. That is, it was formed right before the metallicity dropped, and the exact position of the metallicity dip is sensitive to the accuracy of its age and metallicity. \cite{Mighell1998} used shallower HST photometry and derived an age of 4.7-6.0 Gyr with a metallicity of $-1.16$. A deeper and higher quality HST CMD was analysed by \citet{Glatt2008}, who derived an older age of 6.5 - 7.2 Gyr (the best-fit being with 6.5 Gyr). They assumed a metallicity of $-1.08$ from the previous spectroscopic analysis by \cite{Dacosta1998} in the ZW84 scale. These results supersede those from \cite{Mighell1998}. More recently, \citet{Parisi2015} derived a spectroscopic metallicity of -0.85, consistent with the metallicity from \cite{Dacosta1998}, in the same CG97 scale (i.e. -0.98). These results place K3 in a well-defined position in the AMR, marking the highest metallicity reached in the WH before the metallicity dip.

The other three clusters with ages between 7 and 9 Gyr that define the early chemical evolution together with NGC\,121 are Lindsay\,1, 4, and 6 (see Fig. \ref{Imgamr}). The ages for these clusters were derived from HST and VLT data, whereas metallicities come from CaT analysis by Parisi et al. Therefore, they should be in the same scale as our sample, as discussed in the previous section.

Two clusters from the literature (HW\,6 and Lindsay\,3) join Lindsay\,2, HW\,5, and AM\,3 in defining the metallicity dip. Lindsay\,3 had its age derived from VLT CMD and a spectroscopic metallicity, and its position should be correct with the others. The cluster HW\,6 was only analysed by \cite{Dias+16} with shallower photometry, and therefore the uncertainties are larger. We also note that this cluster is categorised as type `CA' by \citet{Bica+20}, indicating clusters with the property of associations (as in the case of HW\,1 and HW\,5 analysed in this work).

The last two clusters that essentially follow the trend of all clusters younger than 3-4Gyr are Lindsay\,3 and 7. The age of Lindsay\,3 comes from shallower photometry,  whereas VLT data were used for Lindsay\,7. The metallicities come from CaT analysis. Therefore, their positions in the AMR were produced from a consistent analysis.
 





\subsection{Chemical evolution models}

\citet{Tsujimoto&Bekki_2009} discussed the observational evidence of a metallicity dip of $\sim$0.3\,dex around 7.5\,Gyr ago using a compilation of available ages and metallicities for as many SMC clusters as possible that were available at that time. They developed a chemical evolution model to describe this AMR and concluded that the most likely explanation would be an infall of metal-poor gas (with mass $\sim10^8 M_{\odot}$) during a major merger. Since then, many more works have improved the observational AMR of the SMC, showing an intrinsically large dispersion in metallicity and dissolving any strong evidence of a metallicity dip \citep[e.g.][]{Parisi2009,2011Piatti,Parisi2015,PapI,PapIII,PapIV,parisi+22,deBortili2022,milone+22}. Nevertheless, \citet{Dias+16} have shown evidence that this metallicity dispersion is significantly reduced when the WH is analysed separately, claiming that different SMC regions may have had different chemical evolution histories. \citet{Dias+19Zenodo} revealed a metallicity dip in the AMR of the SMC WH based on preliminary results using the VISCACHA data that is now confirmed in this work. We ran new models to explain a major merger signature in a specific region of the SMC.

We calculated the time evolution of Fe abundance by incorporating the major merger process into the model to deduce the age $-$ [Fe/H] relation. We adopted the same merger model of \citet{Tsujimoto&Bekki_2009}; however, some updates were considered and introduced in this study. First, the understanding of delay time distribution (DTD) of type Ia supernovae (SNe Ia) has advanced in recent decades from the observation of external galaxies by extensive supernova surveys. According to the knowledge thus obtained, we assumed DTD $\propto t_{\rm delay}^{-1}$ with a range of $0.1\leq t_{\rm delay}\leq10$ Gyr \citep{Maoz+14}. Secondly, we considered the cases with an enhanced star formation triggered by either a major merger or an external disturbance owing to tidal interaction with the LMC \citep[e.g.][]{Bekki_Chibba+05}. For the initial mass function (IMF), we assumed a power-law mass spectrum with a slope of $-1.35$ (i.e. the Salpeter IMF). The star formation rate (SFR) was assumed to be proportional to the gas fraction with a constant coefficient of $\nu$, which is defined as a reciprocal of a timescale of star formation.
The overall chemical evolution of the SMC is characterised by a slow star formation. The rate coefficient of star formation ($\nu$)  is assumed to be 0.03 Gyr$^{-1}$, which is smaller by more than one order of magnitude than the value in the solar neighbourhood. An enhanced star formation is given by $\nu = $ 0.05 Gyr$^{-1}$.
Both quantities are in agreement with the findings by \citet{nidever+20}. The major merger process was modelled by two parameters: (1)~the mass ratio between the original gas belonging to the specific region of the SMC and the accreted gas onto it and (2)~the time of the merger. To reproduce the observed feature of a dip in [Fe/H], we assumed a mass ratio of 1:4 with the onset of the merger at 6 Gyr ago (green curve in Fig. \ref{Imgamr}). This means that the major event triggers an increase in the SFR of one order of magnitude in the models. In addition, we considered an alternative scenario where the merger did not enhance the SFR, but the SFR was enhanced with a value of $\nu = $ 0.08 Gyr$^{-1}$ in a more recent event about $\sim4$~Gyr ago (red curve in Fig. \ref{Imgamr}).
The resultant age-[Fe/H] relations compared to the clusters' values are shown in Figure \ref{Imgamr}. The two models represent well the average behaviour of the WH star clusters, with a preference for the green line representing the merger model causing the enhancement in the SFR. Also in  Fig. \ref{Imgamr}, we display two extra models that were used to test some limits in the SFR enhancement and mass ratio. A model with the same 1:4 mass ratio but no SFR enhancement wraps the data in the metal-poor boundary of the AMR. The other model has a 1:2 mass ratio and an enhanced SFR ($\nu = $ 0.065 Gyr$^{-1}$) and wraps the data in the metal-rich boundary of the AMR. These two models would represent the limiting scenarios in the predictions.
This supports the major merger scenario that would involve the complex processes of an internal mixing of gas and an induced star formation. Additionally, the Fe enrichment associated with the dynamical effect by the SMC-LMC interaction could help better explain the $\sim0.2$~dex scatter in the observed relation. \\
\begin{figure*}[t]
    \centering
      \begin{subfigure}[b]{\textwidth}
         \includegraphics[width=\textwidth]{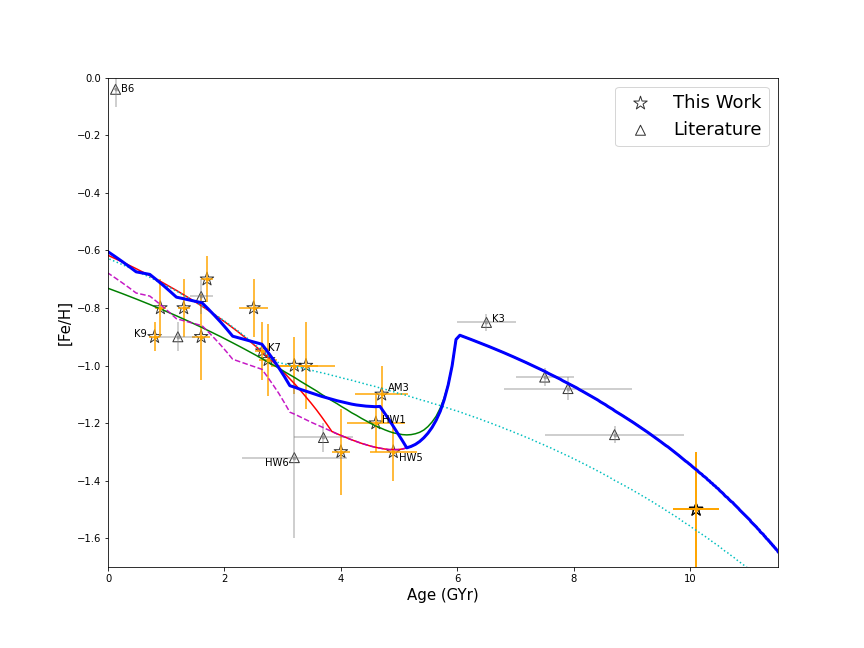}
     \end{subfigure}
    \caption{ The points, the green and red curves are the same as in Figure \ref{Imgamr}. The cyan dotted curve shows the case with no merger and a burst of star formation 3~Gyr ago. The magenta dashed curve represents the scenario with multiple bursts at $3$~Gyr, $2$~Gyr, $1.1$~Gyr, and $0.45$~Gyr with a duration of 0.5~Gyr each (\citealp{2022Massana}). The blue solid curve illustrates the aforementioned scenario of multiple bursts with an additional burst 5~Gyr ago (\citealp{2015RUBELE}).  }
    \label{Newmodels}
\end{figure*}    
Apart from the simple models that we discussed above to describe our cluster data, we added some independent results from other works with field stars to check whether they agree with the star clusters.
Previous work in the literature about the star formation history (SFH) of the SMC identified the presence of enhanced star formations within the last 5 Gyr. Using the deep photometry from the SMASH survey, \citet{2022Massana} identified distinctive peaks of star formations at around $3$~Gyr, $2$~Gyr, $1.1$~Gyr, and $0.45$~Gyr. In Fig. \ref{Newmodels}, we present a model with the merger followed by this sequence of bursts with a duration of 0.5 Gyr for each with the same SFR of 0.1 Gyr$^{-1}$. The model (magenta line) appears to be consistent with the data and with our preferred simple models (red and green curves). The previous study by
\citet{2015RUBELE} using the VMC survey detected periods of enhanced star formation at ages of 1.5 and 5 Gyr ago. Another model was generated by including this older burst of 5~Gyr with an SFR of 0.015 Gyr$^{-1}$ to Massana's sequence of bursts along with a major merger event. This model is also shown in Fig. \ref{Newmodels} (blue line), and it is also consistent with the data and in good agreement with the simple models described before.
We also considered the very popular chemical evolution burst model by \citet{PT98}, who proposed a null SFR for a quiescent period followed by a burst 3 Gyr ago. The assumption of a null SFR is no longer valid based on recent observations by \citet{rubele+18,2022Massana}, for example. Therefore, we performed an exercise where we made a model that follows a constant SFR of 0.015 Gyr$^{-1}$ combined with a burst 3~Gyr ago increasing the SFR to 0.065 Gyr$^{-1}$ and no merger involved. This model is represented by the cyan line in Fig. \ref{Newmodels}, and it does not reproduce the data by any means.

\subsection{ Major merger and implications on the chemical enrichment}

The model that best describes the observed clusters' ages and metallicities is a major merger with an infall of metal-poor gas 6~Gyr ago that was able to speed up the SFR locally (green curve in Fig. \ref{Imgamr}).  
In \cite{parisi+22}, we saw that the most probable model for the WH was the closed box model of DH98, with the exception of two points. It is possible that the 6~Gyr dip was present there but not clearly visible due to the lack of clusters between 4~Gyr and 6~Gyr in that sample. The present work fills in that age interval, making the dip in metallicity more evident.
Moreover, this evidence seems not to be present when the entire SMC is analysed \citep[e.g.][]{Dias+16,parisi+22}.
This necessarily implies the need of a history of inefficient gas mixing within the entire SMC  combined with different SFR at different regions, as supported by other evidence from spectroscopy \citep[e.g.][]{parisi+22,debortoli+22,Mucciarelli2023} and kinematics \citep[e.g.][]{zivick+2021, piatti+21, Dias+22, ElYoussoufi2023}.

The merging of a gas-rich dwarf with the SMC several Gyr ago could have destroyed the original (first generation) thin disc of the SMC, and the old disc would have become a spheroidal component. However, the gas components from the SMC and the dwarf would have formed a new gas disc from which a second generation stellar disc could have formed \citep[e.g.][]{Bekki+Chiba2008}. Thus, the present disc of the SMC would be dominated by the second generation disc composed of new stars formed after a previous merger event. Indeed, old components (older than several Gyr) of the SMC do not clearly do show rotational kinematics (\citealp{HarrisZaritsky2006}), whereas younger stars and gas clearly show rotation (\citealp{Stanimirovic+2004}).

\citet{rubele+18} analysed SMC field stars mainly from the main body and the SMC wing, and they found an AMR with an early chemical enrichment followed by an almost constant metallicity around -0.5~dex from $\sim2$~Gyr ago until recently with no evidence of a metallicity dip. This difference could be related to the different SMC regions analysed, in agreement with other works based on star clusters \citep[e.g.][]{Dias+16,parisi+22}, or it could be related to the differences between star cluster and field star formation. It might be possible that clusters could only form from infalling metal-poor gas that collided with the SMC gas disc, whereas the field stars could have formed everywhere in the SMC gas disc before it was destroyed by the merging.

Moreover, we note that the recent work by Oliveira et al. (2023; VISCACHA paper VII) shows a metallicity dip about 1.5-2 Gyr ago based on the analysis of SMC clusters only from the WB region. This time coincides with the LMC-SMC-MW interaction that caused the Magellanic stream formation, which would be enough to produce a metallicity dip and star formation burst without requiring a merger to explain it. Oliveira et al. have also argued that the clusters younger than about 200 Myr were formed in situ from the gas removed from the SMC inner disc that was enriched to about $-0.5$ dex. If we put this scenario together with the one discussed above, we could say that there was a major merger 6~Gyr ago that formed the WH, destroyed the SMC disc, formed a second generation disc that was chemically enriched by stellar evolution, and served as the material that  was retrieved to the opposite side of the WH to form the the Magellanic Bridge. This speculative scenario must be deeply analysed by chemo-dynamical models in the future using this new observational evidence before it can be confirmed or refuted.  For example, [Mg/Fe] shows a very small spread ($\lesssim 0.1$ dex) for the merger model related to metallicity dip that requires very high-resolution spectroscopy to be detected. This evidence was shown and discussed by \cite{Tsujimoto&Bekki_2009}, and our new models show a very similar pattern.


\section{Summary and conclusions}
In this work, we analysed the $V$, $V-I$ CMDs of 15 of the WH clusters in the SMC based on deep and resolved photometry from the VISCACHA survey. We explored a method of decontamination using a statistical comparison of star samples from the cluster region and an offset field, following the methods in \cite{Maia2010}. The results of the statistical decontamination method show a significant reduction of field star contamination in the CMDs of the sample clusters, allowing for more accurate determination of their physical properties. Overall, the statistical decontamination method is a valuable tool for studying the star clusters, in particular for the mid- to low-mass clusters, and can provide more reliable estimates of their astrophysical properties. The age, metallicity, distance, and reddening of the clusters were estimated by fitting PARSEC isochrones to the decontaminated CMDs. These fits were also corroborated by a good match between isochrones and data from other surveys with different filters covering a wide wavelength range. This agreement gives extra strength to our results.

We added ten other WH clusters with reliable ages and metallicities from the literature to make a combined sample. We checked if there were any systematic differences between the derived ages and metallicities in this work and those from the literature using the clusters the two samples have in common, and we found no significant offset. We found the astrophysical properties of the clusters from the previous studies to be consistent with our analysis, validating the effectiveness of the decontamination method. Our metallicity estimates also show reasonable agreement with most of the spectroscopic metallicities from the literature.

Through this work, we derived the most up-to-date AMR of the SMC WH clusters in order to constrain one piece of the SMC chemical evolution puzzle. We found a low dispersion AMR revealing a metallicity dip of about 0.5 dex around 6~Gyr ago. We ran chemical evolution models, and we found the scenario that best explains this AMR is a major merger with a metal-poor gas cloud or dwarf galaxy with a mass of about $10^8 M_{\odot}$ (mass ratio 1:4) that also enhanced the SFR of the SMC WH. This scenario requires the SMC to not have been very efficient at mixing its gas throughout its life and for the SFR to be different at different SMC regions because the AMR of star clusters from the entire SMC together does not show a metallicity dip \citep[e.g.][]{parisi+22} and neither do the field stars from the main body nor the SMC wing \citep{rubele+18}. Even though \cite{parisi+22} analysed the WH separately, they could not observe the dip due to sample incompleteness between 4 and 6~Gyr.

We speculate that this new observational evidence, combined with the fact that the WH is moving away from the SMC (e.g. \citealp{niedefhofer+18,zivick+18,PapIV}), could be explained by a scenario wherein the SMC suffered a major merger 6~Gyr ago. The merger 
in this scenario destroyed SMC's thin disc, enhanced the local SFR, and formed a new disc from which the material to form the Magellanic Bridge was taken. This scenario is in line with previous simulations from \citet{Bekki+Chiba2008} and \citet{Tsujimoto&Bekki_2009}.
Full chemo-dynamical models should be updated against these new observational results to confirm or refute this proposed scenario. The detailed analysis of star clusters in other SMC regions, as done in this work and by \citealp[]{2023Raphael} on the SMC bridge, should be completed in order to provide a full set of constraints for the chemo-dynamical models.

\section*{Acknowledgements}
S.S. and B.D. acknowledge support by ANID-FONDECYT iniciación grant No. 11221366. 
D.M. gratefully acknowledges the support by the ANID BASAL projects ACE210002 and FB210003 and by Fondecyt Project No. 1220724. \textbf{DM Also acknowledges support from CNPq/Brazil through project350104/2022-0.}
T. T. acknowledges the support by JSPS KAKENHI Grant Nos. 18H01258, 19H05811, and 23H00132.
F. Maia acknowledges financial support from Conselho Nacional de Desenvolvimento Científico e Tecnológico - CNPq (proc.404482/2021-0) and from FAPERJ (proc. E-26/201.386/2022 and E-26/211.475/2021).
R.A.P.O. acknowledges FAPESP support (proc. 2018/22181-0). 
MCP acknowledge support by the Argentinian institutions CONICET (Consejo Nacional de Investigaciones Científicas y Técnicas), SECYT (Universidad Nacional de Córdoba) and ANPCyT (Agencia Nacional de Promoción Científica y Tecnológica).
\textbf{L.O.K. acknowledges partial financial support by CNPq (proc. 313843/2021-0) and UESC (proc. 073.6766.2019.0013905-48).}
Based on observations obtained at the Southern Astrophysical Research (SOAR) telescope (projects SO2016B-018, SO2017B-014, CN2018B-012, SO2019B-019, SO2020B-019, SO2021B-017), which is a joint project of the Ministério da Ciência, Tecnologia, e Inovação (MCTI) da República Federativa do Brasil, the U.S. National Optical Astronomy Observatory (NOAO), the University of North Carolina at Chapel Hill (UNC), and Michigan State University (MSU).
This work has made use of data from the European Space Agency (ESA) mission {\it Gaia} (\url{https://www.cosmos.esa.int/gaia}), processed by the {\it Gaia} Data Processing and Analysis Consortium (DPAC, \url{https://www.cosmos.esa.int/web/gaia/dpac/consortium}). Funding for the DPAC has been provided by national institutions, in particular the institutions participating in the {\it Gaia} Multilateral Agreement.
This research uses services or data provided by the Astro Data Lab at NSF’s NOIRLab. NOIRLab is operated by the Association of Universities for Research in Astronomy (AURA), Inc. under a cooperative agreement with the National Science Foundation.
Based on data products created from observations collected at the European Organisation for
Astronomical Research in the Southern Hemisphere under ESO programme 179.B-2003.

\bibliography{biblio.bib}

\appendix
\section{Extra material}

\begin{table}[H]
    \centering
    \caption{Observation log for the sample clusters.}
    \label{tablog}
    \begin{tabular}{p{2cm}p{2cm}p{2cm}p{2.8cm}p{1cm}p{2.5cm}p{1.5cm}p{1.5cm}}
    \hline
        \noalign{\smallskip}
        Cluster Name & RA & Dec& Date & Filter& Exposure time& Airmass&FWHM\\
         & [h:m:s] & $[^\circ:':"]$ & [YYYY-MM-DD]& & sec &  &   \\ 
         \noalign{\smallskip}
         \hline\hline
         \noalign{\smallskip}
         L2     &00:12:55.00    &-73:29:12.0   &2019-10-04   &V, I  &$3\times400$, $3\times600$ & 1.61, 1.56  & 0.81, 0.74  \\
         HW1    &00:18:27.00    &-73:23:42.0   &2021-11-07   &V, I  &$3\times400$, $3\times600$ & 1.40, 1.38  & 0.70, 0.62  \\
         B2     &00:19:18.00    &-74:34:30.0   &2021-11-09   &V, I  &$3\times400$, $3\times600$ & 1.42, 1.41  & 0.86, 0.70  \\
         B1     &00:19:20.00    &-74:06:24.0   &2016-11-02   &V, I  &$6\times200$, $6\times300$ & 1.42, 1.41  & 0.71, 0.64  \\
         B4     &00:24:54.00    &-73:01:48.0   &2021-11-08   &V, I  &$4\times100$, $6\times100$ & 1.39, 1.38  & 0.62, 0.49  \\
         K6     &00:25:26.60    &-74:04:29.7   &2016-11-03   &V, I  &$6\times200$, $6\times300$ & 1.43, 1.41  & 1.08, 0.91  \\
         NGC121 &00:26:49.00    &-71:32:10.0   &2017-10-19   &V, I  &$4\times200$, $4\times300$ & 1.33, 1.33  & 0.56, 0.45  \\
         K7     &00:27:45.17    &-72:46:52.5   &2016-09-28   &V, I  &$6\times200$, $6\times300$ & 1.38, 1.37  & 0.64, 0.49  \\
         K8     &00:28:02.14    &-73:18:13.6   &2017-10-20   &V, I  &$4\times200$, $4\times300$ & 1.46, 1.44  & 1.37, 1.56  \\
         K9     &00:30:00.26    &-73:22:40.7   &2016-11-04   &V, I  &$6\times200$, $6\times300$ & 1.39, 1.41  & 0.96, 0.99  \\
         HW5    &00:31:01.34    &-72:20:30.0   &2016-11-05   &V, I  &$3\times200$, $3\times300$ & 1.74, 1.77  & 0.85, 0.86  \\
         L14    &00:32:41.02    &-72:34:50.1   &2021-11-07   &V, I  &$3\times400$, $3\times600$ & 1.36, 1.36  & 0.63, 0.56  \\
         NGC152 &00:32:56.26    &-73:06:56.6   &2016-11-05   &V, I  &$4\times200$, $4\times300$ & 1.37, 1.37  & 0.71, 0.45  \\
         K11    &00:36:26.62    &-72:28:43.7   &2017-10-20   &V, I  &$4\times200$, $4\times300$ & 1.36, 1.36  & 1.06, 1.09  \\
         AM3    &23:48:59.00    &-72:56:42.0   &2016-11-05   &V, I  &$6\times200$, $6\times300$ & 1.38, 1.37  & 0.51, 0.38  \\ 
         \noalign{\smallskip}
         \hline
    \end{tabular}
\end{table}

\newpage
\begin{figure*}
    \centering

    \begin{subfigure}[b]{\textwidth}
         \includegraphics[width=0.98\textwidth]{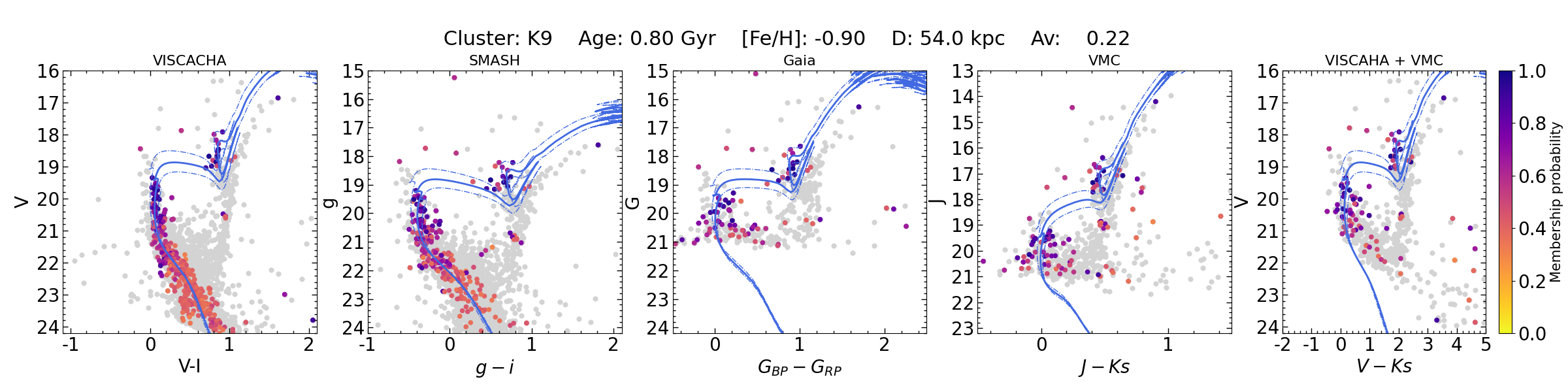}
     \end{subfigure}

      \begin{subfigure}[b]{\textwidth}
         \includegraphics[width=0.98\textwidth]{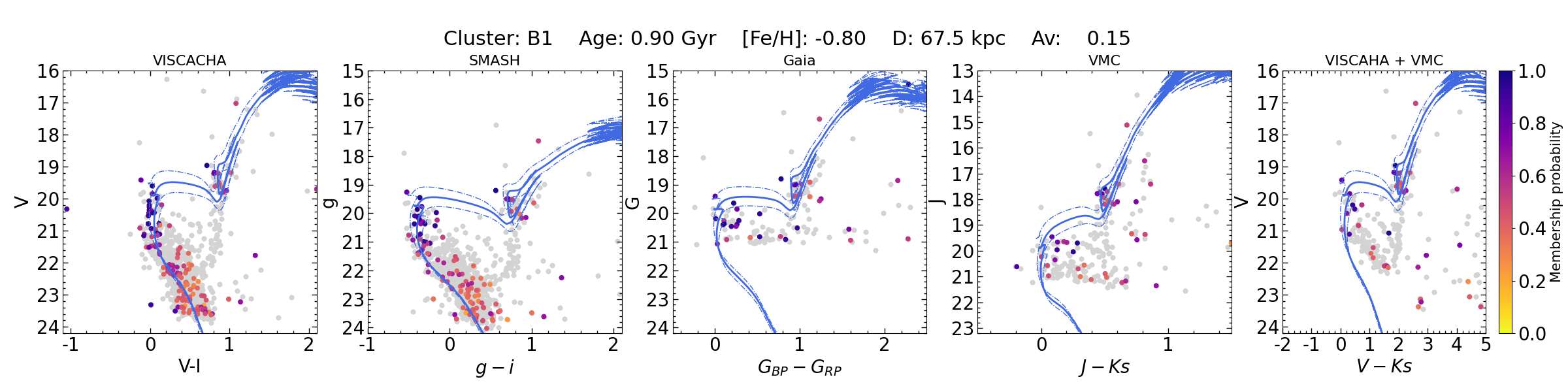}
     \end{subfigure}

     \begin{subfigure}[b]{\textwidth}
         \includegraphics[width=0.98\textwidth]{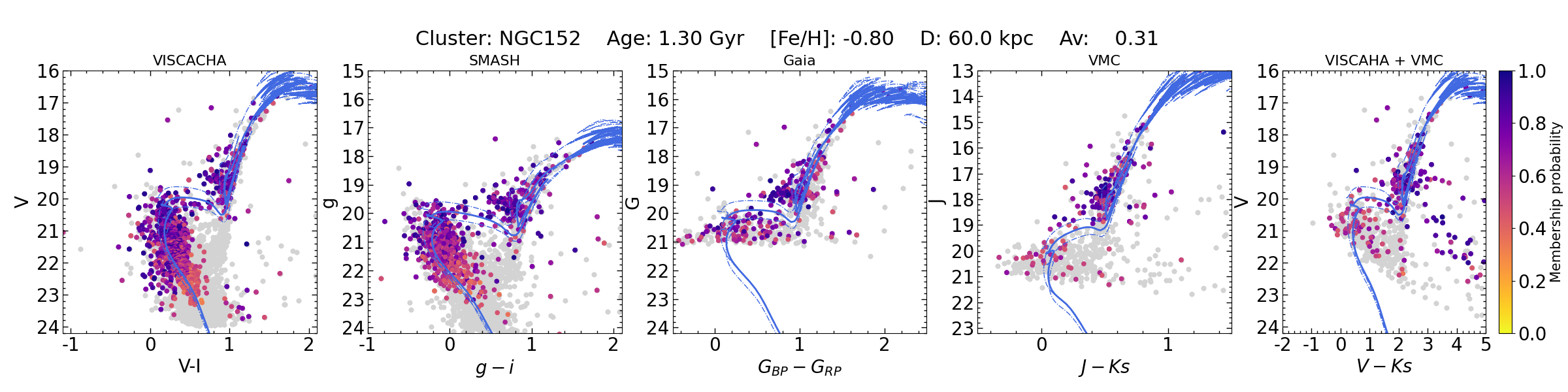}
     \end{subfigure}

      \begin{subfigure}[b]{\textwidth}
         \includegraphics[width=0.98\textwidth]{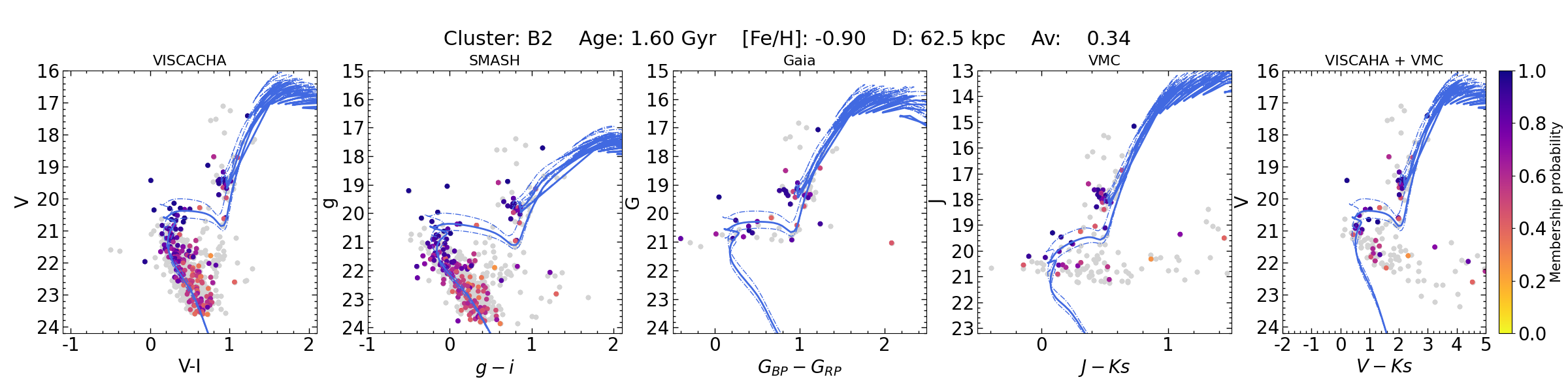}
     \end{subfigure}

    \begin{subfigure}[b]{\textwidth}
        \includegraphics[width=0.98\textwidth]{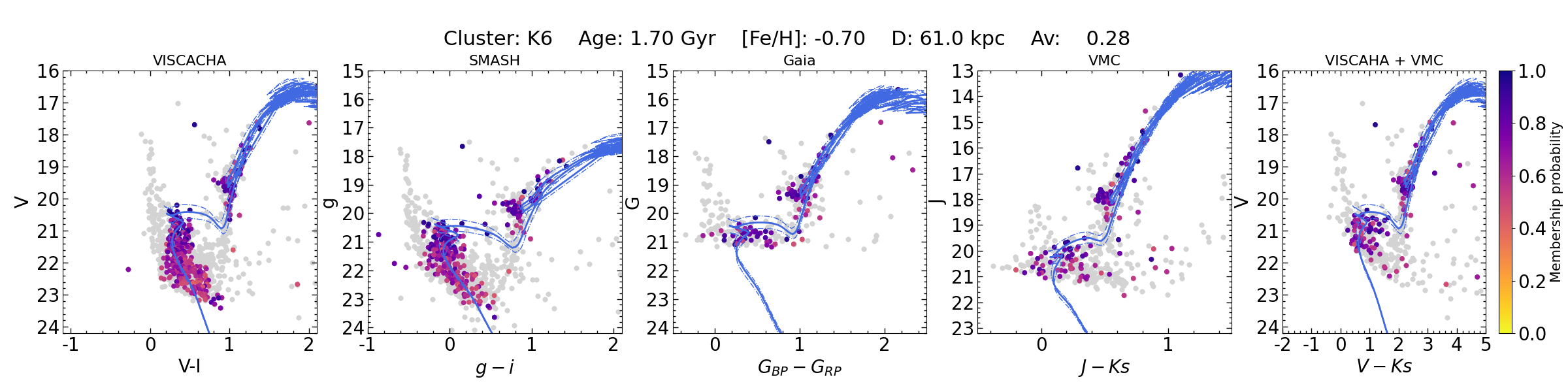}
    \end{subfigure}  
    \caption{ Best fitting isochrones for the decontaminated CMDs (similar to the ones shown in Figure \ref{ImgdecontCMD}) of several sample clusters. The CMDs for each cluster from different surveys, including SMASH, Gaia, and VMC, are plotted in their respective filters and labelled accordingly.
    The membership probability derived for the VISCACHA sample was adopted for the matched stars in other catalogues as well. The solid line represents the best fitting isochrone, while the dash-dotted isochrones indicate the boundaries that determine the uncertainties of the derived parameters.  }
    \label{Imgtrial01}
\end{figure*}
\begin{figure*}
    \centering

    \begin{subfigure}[b]{\textwidth}
         \includegraphics[width=\textwidth]{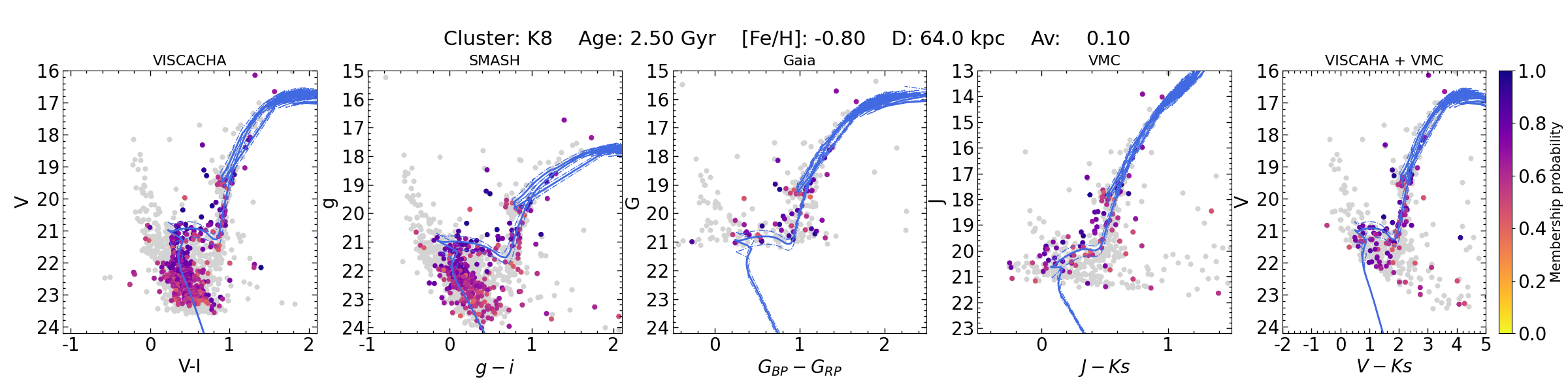}
     \end{subfigure}

      \begin{subfigure}[b]{\textwidth}
         \includegraphics[width=\textwidth]{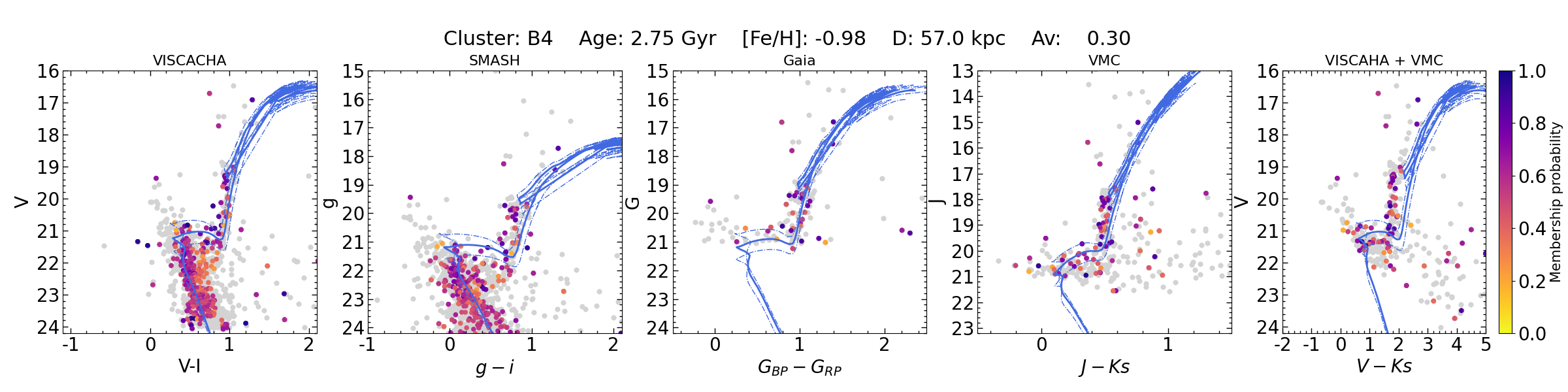}
     \end{subfigure}

     \begin{subfigure}[b]{\textwidth}
         \includegraphics[width=\textwidth]{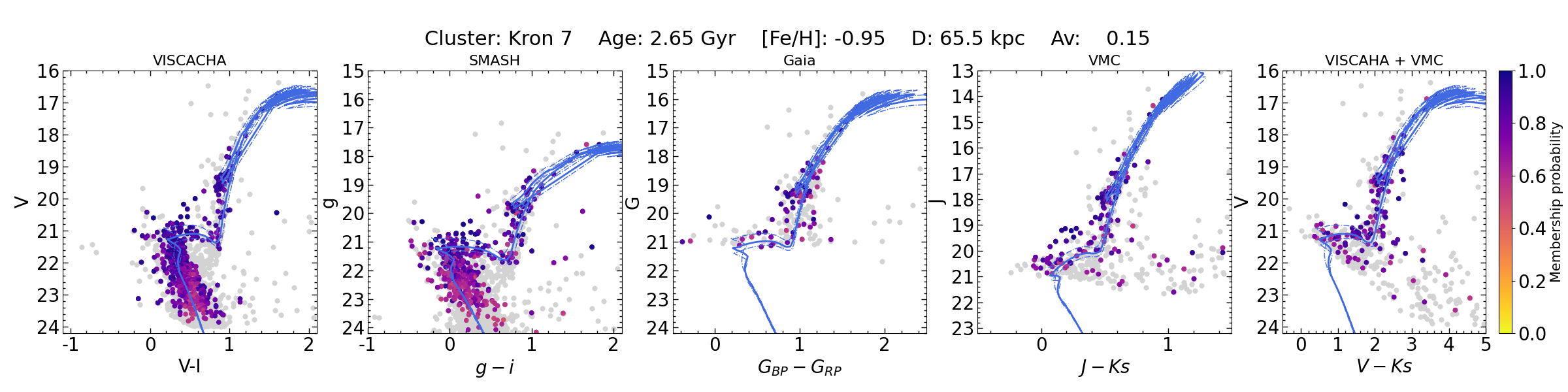}
     \end{subfigure}

      \begin{subfigure}[b]{\textwidth}
         \includegraphics[width=\textwidth]{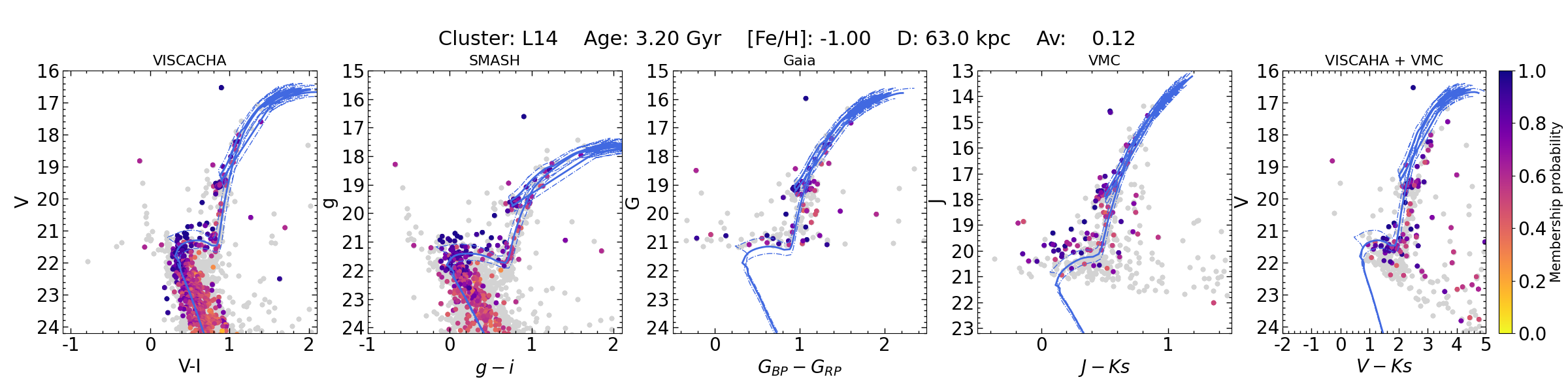}
     \end{subfigure}

    \begin{subfigure}[b]{\textwidth}
        \includegraphics[width=\textwidth]{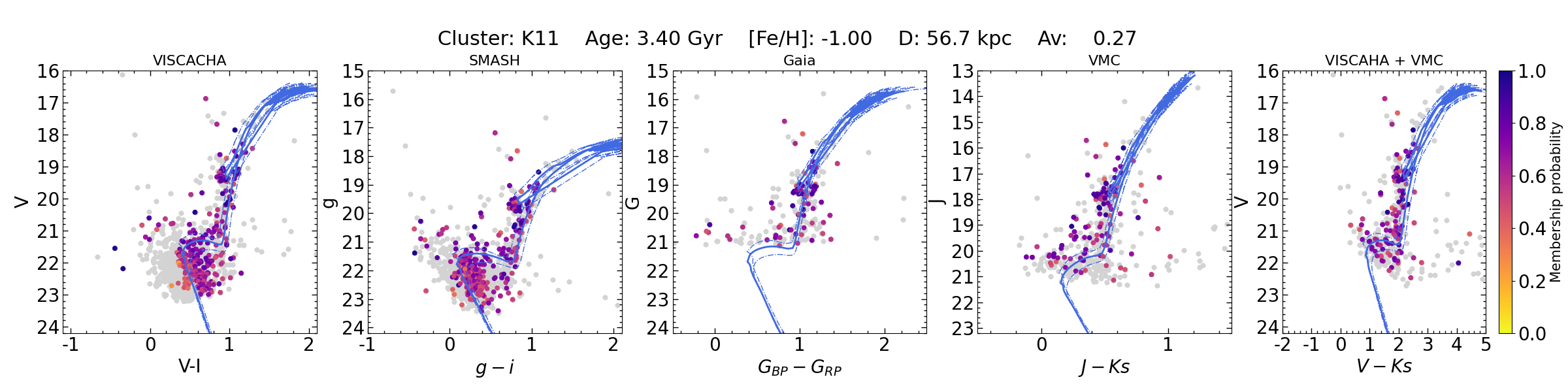}
    \end{subfigure}  
    \caption{Same as Fig. \ref{Imgtrial01} but for more clusters.}
    \label{Imgtrial02}
\end{figure*}
\begin{figure*}
    \centering
      \begin{subfigure}[b]{\textwidth}
         \includegraphics[width=\textwidth]{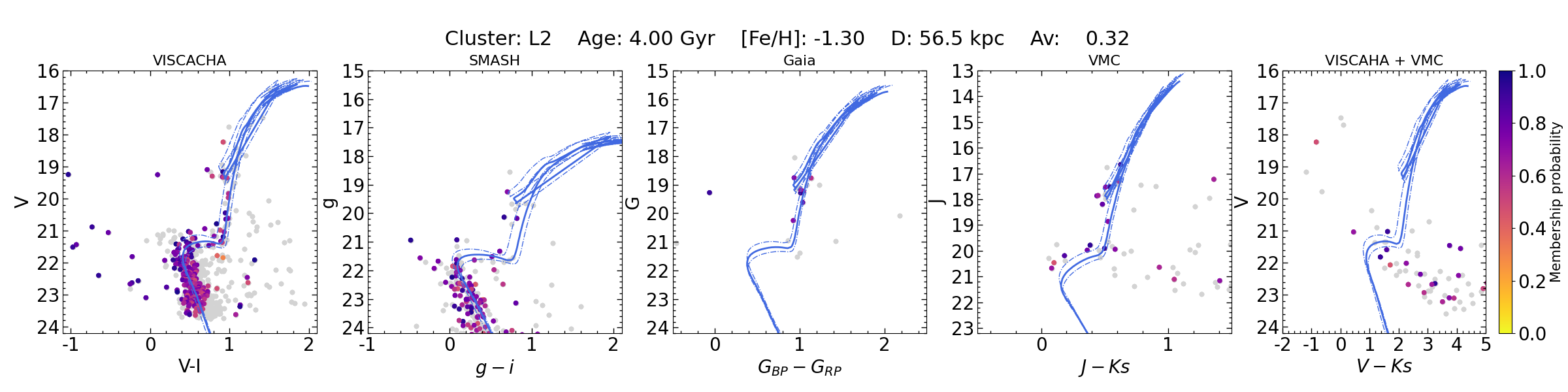}
     \end{subfigure}    
    \begin{subfigure}[b]{\textwidth}
         \includegraphics[width=\textwidth]{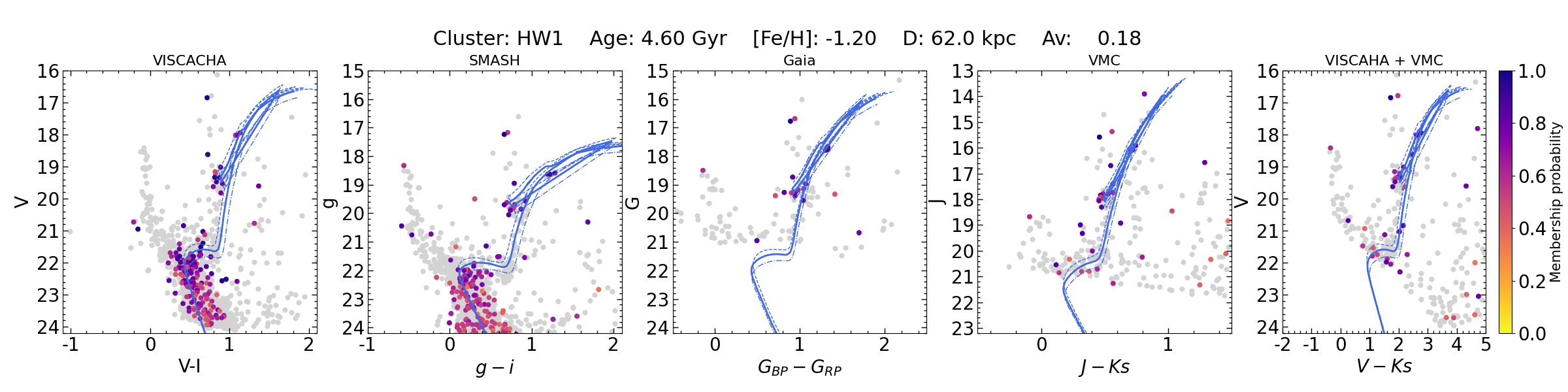}
     \end{subfigure} 
      \begin{subfigure}[b]{\textwidth}
         \includegraphics[width=\textwidth]{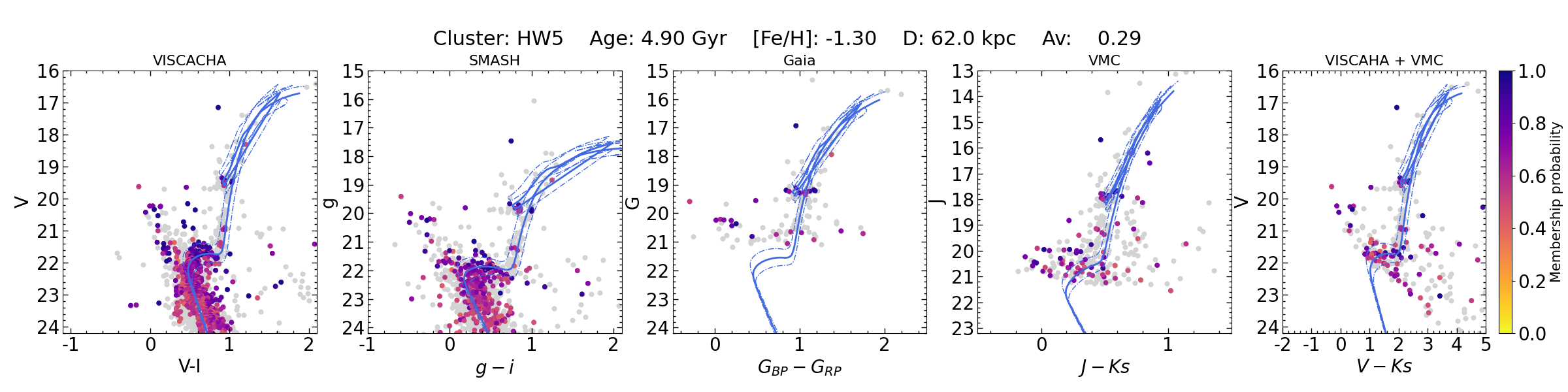}
     \end{subfigure}
     \begin{subfigure}[b]{\textwidth}
         \includegraphics[width=\textwidth]{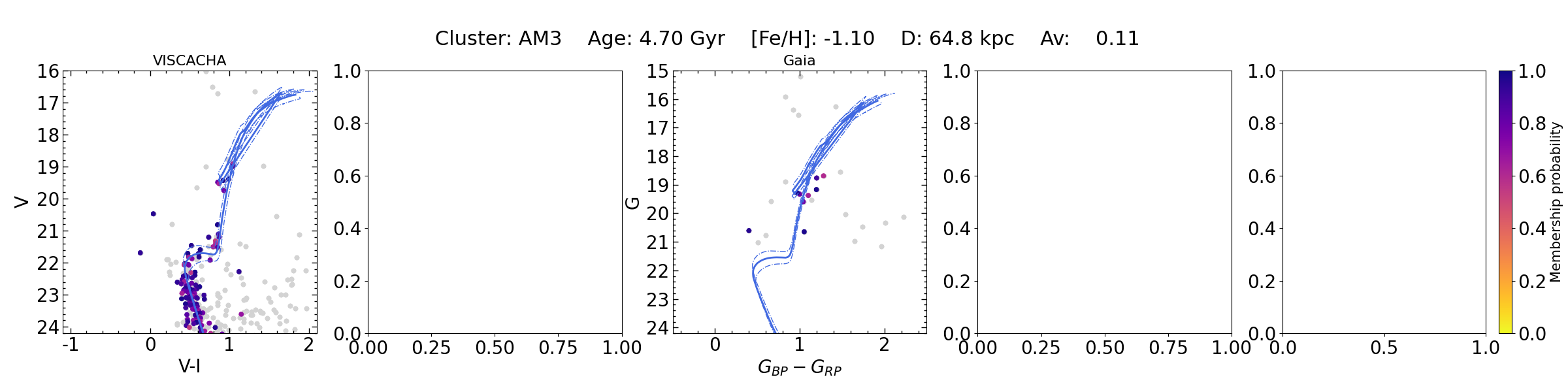}
     \end{subfigure}
    \begin{subfigure}[b]{\textwidth}
        \includegraphics[width=\textwidth]{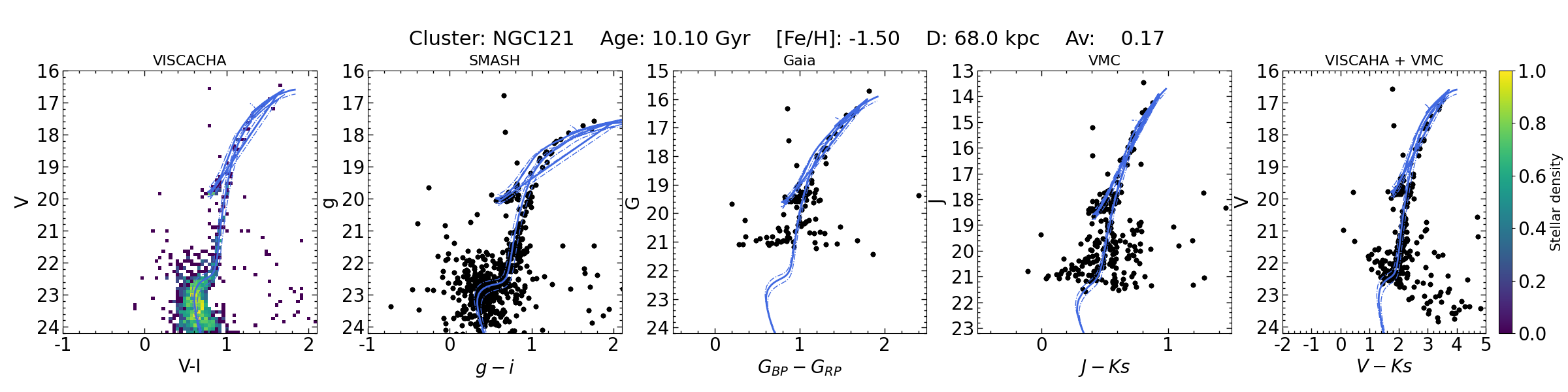}
    \end{subfigure}  
    \caption{Same as Fig. \ref{Imgtrial01} but for more clusters. The bottom panels represent NGC 121 and have a different colour code (explanation is given in Section 4). The colours represent the normalised stellar number density of the stars in the CMD and clearly show that the isochrone is tracing well-defined CMD features. As the density of stars is lower in other surveys, the colour code was not adopted for them.}
    \label{Imgtrial03}
\end{figure*}
\begin{table*}
    \centering
    \footnotesize
    \caption{Age and metallicity estimates of all the west halo clusters available from the literature.}
    \label{tabcomp}
    \begin{tabular}{p{2cm}p{2cm}p{2cm}p{2cm}p{2cm}}
    \hline
        \noalign{\smallskip}
         Cluster & Age & Reference & [Fe/H] &Reference \\
                 &[Gyr]&           &        &           \\
         \hline\hline
         
         \multicolumn{5}{c}{WH clusters already studied in common with our sample}\\
         
         \hline
         \noalign{\smallskip}
         K9, L13     &0.40                      &1          & $-1.12\pm0.05$                  &2              \\
                      &0.50                      &3          &  $-$                           &  $-$            \\
         \noalign{\smallskip}
         NGC152      &$1.27_{-0.26}^{+0.07}$      &4           & $-0.77_{-0.21}^{+0.07}$        &4               \\
                \noalign{\smallskip}
                      &$-$      & $-$           &$-0.72 \pm 0.02$         &5               \\
                      
                      &$-$      & $- $          &$-0.73 \pm 0.11$         &6               \\
                    
                     &$-$      & $-$            &$-1.25 \pm 0.25$         &7               \\
                   
                       &$1.40\pm0.20$               &8           &$-0.94\pm0.15 $                &8                \\
                      
                        &$1.23\pm0.07$               &9           &$-0.87\pm0.07$                &9                 \\
         \noalign{\smallskip}
         B2          &$1.80\pm0.70 $              &9           &$-1.00\pm $                     &9               \\
         \noalign{\smallskip}
         K6, L9      &$1.60\pm0.20$               &10           &$-0.63\pm0.03$                  &2              \\
         \noalign{\smallskip}
         K8, L12     &$2.15_{-0.21}^{+0.21}$      &4            &$-0.75_{-0.07}^{+0.07}$        &4               \\
         \noalign{\smallskip}
                      &1.30                      &11           &$-0.70\pm0.03$                &2              \\
                      \noalign{\smallskip}
                       &$2.94\pm0.31$               &9            &$-1.12\pm0.15$               &9               \\
         \noalign{\smallskip}
         B4          &$3.80\pm0.60$               &9           &$-1.19\pm0.24$                  &9               \\
         \noalign{\smallskip}
         K7          &$2.34_{-0.08}^{+0.20}$      &4           &$-1.04_{-0.05}^{+0.05}$         &4               \\
         \noalign{\smallskip}
                      &$3.50\pm1.00$               &12,10         &$-0.81\pm0.04$                 &12                \\
                       &3                         &13           &$-0.80\pm0.04$                &13                 \\
         \noalign{\smallskip}
         L14         &$2.80\pm0.40$               &9           &$-1.14\pm0.11$                  &9               \\
         \noalign{\smallskip}
         K11         &$1.47\pm0.11$               &9           &$-0.78\pm0.19$                  &9               \\
         \noalign{\smallskip}
         L2          &$3.98_{-0.55}^{+0.37}$      &4           &$-1.27_{-0.08}^{+0.10}$         &4               \\
         \noalign{\smallskip}
                      &$4.00_{-0.7}^{+0.9}$        &14           &$-1.40_{-0.2}^{+0.2}$          &14               \\
         \noalign{\smallskip}
         HW1         &$5.00_{-1.2}^{+1.5}$        &14           &$-1.10_{-0.3}^{+0.2}$           &14               \\
         \noalign{\smallskip}
         HW5         &$4.30\pm0.90$               &9           &$-1.28\pm0.32$                  &9               \\
         \noalign{\smallskip}
         AM3         & $5-6$                     &15          & $\sim -$1.30                 & 15               \\
                    & $6.00\pm0.15$              &16           &$-1.25\pm0.25$               &16              \\
         \noalign{\smallskip}
                     &$5.48_{-0.74}^{+0.46}$      &17           &$-1.36_{-0.25}^{+0.31}$        &17               \\
         \noalign{\smallskip}
                     &$4.90_{-1.5}^{+2.1}$        &14           &$-0.80_{-0.6}^{+0.2}$         &14               \\
         \noalign{\smallskip}
                     &$4.40_{-1.4}^{+1.3}$        &4           &$-1.00_{-0.10}^{+0.10}$         &4               \\ 
         \noalign{\smallskip}
         NGC121      &$10.5\pm0.50$               &18           &$-1.19\pm0.12$                  &12               \\
                     &$-$                         &$-$            &$-1.28\pm0.06$                 &19               \\
                     &$-$                         &$-$            &$-1.41\pm0.42$                 &20               \\
                     &$-$                         &$-$            &$-1.51\pm0.42$                 &21               \\   
         \hline       
         \multicolumn{5}{c}{WH clusters already studied, but not in our sample}\\       
         \hline
         \noalign{\smallskip}
         B6          &$0.13\pm0.04$               &9           &$-0.04\pm0.06$                  &9               \\
         \noalign{\smallskip}
         L3          &$1.20_{-0.3}^{+0.3}$        &14           &$-0.40_{-0.1}^{+0.2}$           &14               \\
                      &$-$                         &$-$           &$-0.90\pm0.05$                 &22              \\  
         \noalign{\smallskip}
         K5, L7      &$1.60\pm0.20$               &23          &$-0.76\pm0.06$                  &24              \\             
         \noalign{\smallskip}
         HW6         &$3.20\pm0.90$               &9           &$-1.32\pm0.28$                  &9               \\
         \noalign{\smallskip}
         L5          &$3.70\pm0.50$               &23          &$-1.25\pm0.05$                  &24              \\
         \noalign{\smallskip}
         K3, L8       &$6.50\pm0.50$               &25          &$-0.85\pm0.03$                  &2              \\
                      &$3.5\pm 1$                  &22          &$-0.93$                        &22              \\
                       &$4.70\pm0.60$               &27          &$-1.16\pm0.09 $                &27              \\
         \noalign{\smallskip}
         L1          &$7.50\pm0.50$               &25          &$-1.04\pm0.03$                  &2              \\
         \noalign{\smallskip}
         K1, L4      &$7.90\pm1.10$               &23          &$-1.08\pm0.04$                  &24              \\
         \noalign{\smallskip}
         K4, L6      &$8.70\pm1.20$               &23          &$-1.24\pm0.03$                  &24              \\
         \noalign{\smallskip}
         L11         &$3.00\pm0.40$               &13           &$-0.83\pm0.06$                  &5              \\
         \noalign{\smallskip}
         \hline 
    \end{tabular}
    
      Nearly all clusters from our sample, except B1, were studied before with different techniques. The total 24 clusters plus B1 account for $\sim 60$\% of the total number of WH clusters listed in the \citet{Bica+20} catalogue. The other 16 clusters are: OGLS $263 - 273$, 316, 321, 330, BS4, BS5, H8612, H8615.\\
      
      References : (1) \cite{Nayak2018}; (2) \cite{Parisi2015}; (3) \cite{Glatt2010}; (4) \cite{PapIV}; (5) \cite{parisi+22}; (6) \cite{Song+2021}; (7) \cite{bicaE+1986};  (8) \cite{Crowl2001}; (9) \cite{Dias+16}; (10) \cite{Piatti2005a}; (11) \cite{Rafelski_Zaritsky2005}; (12) \cite{Dacosta1998}; (13) \cite{Livanou2013}; (14) \cite{Dias+14}; (15) \citet{dacosta99}; (16) \cite{2011Piatti}; (17) \cite{PapI}; (18) \cite{Glatt2008}; (19) \cite{Dalessandro16}; (20) \cite{Johnson2004}; (21) \cite{ZinnandWest84}; (22) \cite{deBortili2022}; (23) \cite{Parisi2014}; (24) \cite{Parisi2009}; (25) \cite{Glatt2008b}; (26) \cite{Mould1992}; (27) \cite{Mighell1998}.        
    
\end{table*}

\end{document}